\documentclass[prd,showpacs,preprintnumbers,amsmath,amssymb,showkeys]{revtex4}
\usepackage[dvips]{graphicx,color}% Include figure files
\usepackage{dcolumn}% Align table columns on decimal point
\usepackage{bm}% bold math
  \newcommand{\nn}{\nonumber}

\newcommand{\be}{\begin{equation}}
\newcommand{\ee}{\end{equation}}
\newcommand{\bea}{\begin{eqnarray}}
\newcommand{\eea}{\end{eqnarray}}

\newcommand{\bi}{\begin{itemize}}
\newcommand{\ei}{\end{itemize}}
\newcommand{\benu}{\begin{enumerate}}
\newcommand{\eenu}{\end{enumerate}}

\def\Av{\mbox{\boldmath $A$}}

\def\Cv{\mbox{\boldmath $C$}}

\def\Kv{\mbox{\boldmath $K$}}

\def\Pv{\mbox{\boldmath $P$}}

\def\qv{\mbox{\boldmath $q$}}
\def\kv{\mbox{\boldmath $k$}}

\newcommand\GeV{\ensuremath{\text{GeV}}}
\newcommand\MSbar{\ensuremath{\overline{\text{MS}}}}
\newcommand\DISg{\ensuremath{\text{DIS}_\gamma}}

\begin{document}

%-------   preprint number   ---------%
\preprint{HUPD-0901, YNU-HEPTh-09-102, UTHEP-602, KUNS-2243}

\title{Heavy quark effects on parton
distribution functions in the unpolarized
virtual photon up to the next-to-leading order in QCD}
% Force line breaks with \\

%-------   name of 1st author   -------%
\author{Yoshio Kitadono}
 \email{kitadono@theo.phys.sci.hiroshima-u.ac.jp}
 \affiliation{
 Department of Physics, Faculty of Science,
 Hiroshima University,\\
 Higashi Hiroshima 739-8526, Japan}
 %\altaffiliation[Also]{kitadono@scphys.kyoto-u.ac.jp}

%%-------   name of 2nd author   -------%
\author{Ken Sasaki}
 \email{sasaki@ynu.ac.jp}
 \affiliation{
 Department of Physics, Faculty of Engineering,
 Yokohama National University,\\
 Yokohama, 240-8501, Japan}
%% \altaffiliation[Also at]
%%   { Department of Physics, Fuga-Fuga University.  }
%% Lines break automatically or can be forced with \\
%%---------------------------------------%

%%-------   name of 3rd author   -------%
\author{Takahiro Ueda}
 \email{tueda@het.ph.tsukuba.ac.jp}
 \affiliation{
 Graduate School of Pure and Applied Sciences,
 University of Tsukuba,\\
 Tsukuba, Ibaraki 305-8571, Japan}
%% \altaffiliation[Also at]
%%   { Department of Physics, Fuga-Fuga University.  }
%% Lines break automatically or can be forced with \\
%%---------------------------------------%

%%-------   name of 4th author   -------%
\author{Tsuneo Uematsu}
 \email{uematsu@scphys.kyoto-u.ac.jp}
 \affiliation{
  Department of Physics, Graduate School of Science, Kyoto University,\\
  Yoshida, Kyoto 606-8501, Japan}
% \altaffiliation[Also at]
%%   { Department of Physics, Fuga-Fuga University.  }
%% Lines break automatically or can be forced with \\
%%--------------------------------------%
%
%

%----------------------   date    -----------------------%
\date{\today}% It is always \today, today,
%             %  but any date may be explicitly specified
%--------------------------------------------------------%

%----------------------   abstract    -----------------------%
\begin{abstract}
We investigate the heavy quark mass effects on the parton
distribution functions in the unpolarized virtual photon up to
the next-to-leading order in QCD. Our formalism is based on
the QCD-improved parton model described by the DGLAP evolution
equation as well as on the operator product expansion supplemented
by the mass-independent renormalization group method.
We evaluate the various components of the parton distributions
inside the virtual photon
with the massive quark effects, which are included through
the initial condition for the heavy quark distributions, or
equivalently from the matrix element of the heavy quark operators.
We discuss some features of our results for the heavy quark effects
and their factorization-scheme dependence.
\end{abstract}
%------------------------------------------------------------%

%----   PACS number   ---%

%----   key word   ---%

%---------   title   --------%
\maketitle
%\tableofcontents
%---------------------%

%---------------------------------------------%
\section{Introduction}

The CERN Large Hadron Collider (LHC) has restarted~\cite{LHC} and discoveries of signals for
the new physics beyond the standard model (SM) are much anticipated. In order to fully complement
the discoveries at LHC, we need more precise measurements which will be provided by the
International Linear Collider (ILC), a proposed $e^+ e^-$ collider machine~\cite{ILC}.
In analyzing the signals for the new physics, it is still important for us to have a detailed
knowledge of the SM predictions at high energies based on QCD.

\begin{figure}
\begin{center}
\includegraphics[scale=0.4]{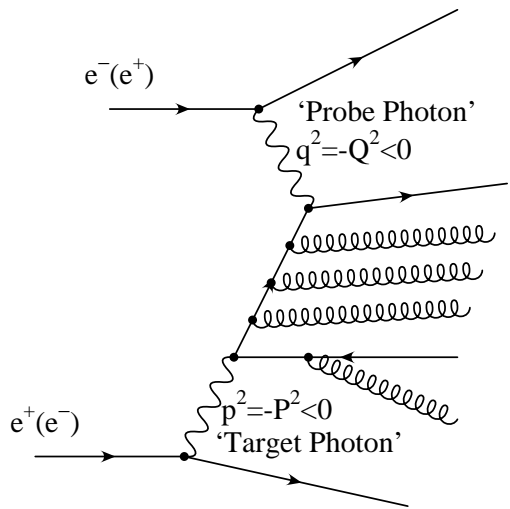}
\caption{Deep inelastic scattering on a virtual photon in the $e^+~e^-$
collider experiments.}
\label{Two Photon Process}
\end{center}
\end{figure}

In $e^+ e^-$ collision experiments at high energies, the cross section of the two-photon processes
$e^+ e^-\rightarrow e^+ e^- + {\rm hadrons}$ dominates over other processes such as the
annihilation processes $e^+ e^-\rightarrow \gamma^* \rightarrow {\rm hadrons}$. The two-photon processes
provide a suitable testing ground for studying the QCD predictions at high energies. We consider here
the two-photon processes in the double-tag events, where both the outgoing $e^+$ and $e^-$ are detected.
In particular, we investigate the case in which one of the virtual photon
is far off-shell (large $Q^2\equiv -q^2$), while the other is close to
the mass-shell (small $P^2=-p^2$). Then the process can be viewed as a
deep-inelastic scattering where the target is a photon rather than a nucleon~%
\cite{twophoton} (see Fig.~\ref{Two Photon Process}).
In this deep-inelastic scattering off photon targets, we can investigate the photon structure functions,
which are the analogues of the nucleon structure functions.
The study of the photon structure functions has long been an active field
of research both theoretically and experimentally~\cite{Review}.

%%%%%%%%%%%%%%%%%%%%%%%%%%%%%%%%%%%%%
%%%%%%%%%%%%%%%%%%%%%%%%%%%%%%%%%%
%%%%%%%

A unique and interesting feature of the photon structure functions is
that, in contrast with the nucleon case, the target mass squared $P^2$
is not fixed but can take various values and that the structure functions show
different behaviors depending on the values of $P^2$.

The unpolarized (spin-averaged) photon structure functions
$F_2^\gamma(x,Q^2)$ and $F_L^\gamma(x,Q^2)$ of the real photon ($P^2=0$) were studied in
the parton model~\cite{QPM}, in perturbative QCD (pQCD) by using
the operator product expansion (OPE)~\cite{CHM} supplemented by the renormalization group (RG)
method~\cite{Witten,BB} and also by using the QCD improved PM~\cite{Altarelli} powered
by the parton evolution equation~\cite{Dewitt,GR1983,MVV2002,MVV2004NNLOpart3}.
The polarized photon structure function
$g_1^\gamma(x,Q^2)$ of the real photon was analyzed in pQCD~%
\cite{polg1LO,polg1NLO1,GRS2001}. The QCD analysis has been made for $F_2^\gamma(x,Q^2)$
up to the next-to-next-to-leading order (NNLO)~\cite{MVV2002} and
for $g_1^\gamma(x,Q^2)$ up to the next-to-leading order (NLO)~\cite{polg1NLO1,GRS2001}.

For a virtual photon target ($P^2\ne 0$), we obtain the
virtual photon structure functions $F_2^\gamma(x,Q^2,P^2)$ and $F_L^\gamma(x,Q^2,P^2)$.
In fact, these structure functions were analyzed in pQCD
for the kinematical region,
\begin{equation}
\Lambda^2 \ll P^2 \ll Q^2~, \label{Kinematical region}
\end{equation}
where $\Lambda$ is the QCD scale parameter~\cite{UW1,UW2,Rossi,BorzumatiSchuler,Chyla}.
The advantage of studying a virtual
photon target in this kinematical region~(\ref{Kinematical region}) is that
we can calculate the whole structure function, its shape and magnitude,
by the perturbative method. This is contrasted with the case of the real photon target
where in the NLO and beyond there appear nonperturbative pieces.
With the recently calculated results of the three-loop anomalous dimensions of
quark and gluon operators~\cite{MVV2004NNLOpart1,MVV2004NNLOpart2}
and also of the three-loop photon-quark and photon-gluon splitting functions~\cite{MVV2004NNLOpart3}
the virtual photon structure function $F_2^\gamma$ ($F_L^\gamma$)
was investigated up to the NNLO (to the NLO) in pQCD~%
\cite{USU2007,KSUU2008}. In the same kinematical region~(\ref{Kinematical region}),
the polarized virtual photon structure function $g_1^\gamma(x,Q^2,P^2)$ was
studied in pQCD~\cite{SU1999,GRS2001,BSU2002,SUU2006}.

In parton picture, structure functions are expressed as
convolutions of coefficient functions and parton distribution functions (PDFs) in the target.
The knowledge of these parton distributions is important since they will be used for predicting the
cross sections of other inclusive processes. When the target is a virtual photon with $P^2$ being
in the kinematical region~(\ref{Kinematical region}),
then a definite prediction can be made for its parton distributions in pQCD.
The parton contents of the unpolarized and polarized virtual photon
for this case were studied in Refs.~\cite{Rossi,DG,GRStratmann,Fontannaz,SU2000}.
Recently the QCD analysis of the parton distributions in the
unpolarized virtual photon target was performed up to the NNLO~\cite{USU2009}.

Although the photon structure functions and photonic parton distributions were studied
in pQCD up to the NNLO~\cite{MVV2002,USU2007,USU2009},
in these analyses all the contributing quarks were assumed to be massless.
The production channel of a heavy flavor (with mass $m$) opens when $(p+q)^2\geq 4m^2$
and its mass effects should be taken into account unless
$Q^2 \gg m^2$. In fact the study of heavy quark mass effects for the two-photon processes
and photon structure functions has appeared in the literature~%
\cite{GR1983,GRV1992b,GRStratmann,Fontannaz,GRS2001,AFG,GRSch1999,SSU2002,CJKL2003,CJK2004}.
Quite recently
the present authors analyzed the heavy quark mass effects on $F_2^\gamma$
for the kinematical region~(\ref{Kinematical region}) up to the NLO~\cite{KSUU2009}
using a different approach from the ones before.
The analysis was made in the framework based on the QCD improved PM and the mass-independent
RG equations, in which the RG equation parameters,
i.e., $\beta$ and $\gamma$ functions, are the same as those for the massless quark
case.

In this paper we examine the heavy quark mass effects on the parton
distribution functions in the unpolarized virtual photon up to
the NLO in pQCD. We use the same framework as the one in Ref.~\cite{KSUU2009},
the QCD improved PM combined with the mass-independent
RG equations. We consider the system which consists of $n_f-1$ light (i.e., massless) quarks and one
heavy quark together with gluons and photons. Then, the heavy quark mass
effects are included in the RG equation inputs; the coefficient functions and the
operator matrix elements.
In the case of the nucleon target, the heavy quark mass effects were studied
by a method based on the OPE in Ref.~\cite{Buza-etal1999}, where the heavy quark
was treated such that it was radiatively generated and absent in the
intrinsic quark components of the nucleon. This picture does not hold
for the case of the photon, since the heavy quark is also generated from the
photon target together with light quarks at high energies. We should consider both the heavy and light quarks
equally as the partonic components inside the virtual photon.

In the next section, we derive the evolution equations for the parton
distribution functions for the case where $n_f-1$ light quarks and one heavy
quark are present. Then solving these equations, we give the explicit expressions up to the NLO
for the moments of the light flavor singlet (nonsinglet) quark, heavy quark
and gluon distributions.
The parton distributions are dependent on the scheme
which is employed to factorize structure functions into
coefficient functions and parton distributions.
We investigate
the photonic parton distributions
in two factorization schemes, namely $\overline {\rm MS}$~\cite{BBDM} and
${\rm DIS}_{\gamma}$~\cite{GRV1992a} schemes.
In Sec.~\ref{Parameters}, we enumerate all the necessary QCD parameters to
evaluate the photonic parton distributions up to the NLO in \MSbar{} scheme.
The parton distributions in \DISg{} scheme are considered in Sec.~\ref{DISg-scheme}.
The numerical analysis of the parton distributions predicted by
$\overline {\rm MS}$ and ${\rm DIS}_{\gamma}$ schemes will
be given in Sec.~\ref{result}.
The final section is devoted to the conclusions.
In Appendix~\ref{relation_massless} we consider the parton distributions in the virtual photon
for the case when all $n_f$ quarks are light.

%%%%%%%%%%%%%%%%
\section{Parton distributions in the virtual photon with a heavy quark flavor}
\label{basic}
%%%%%%%%%%%%%%%%%
We investigate the parton distributions in the virtual photon for the case when one heavy flavor quark
appears together with $n_f - 1$ light (i.e., massless) quarks.
The analysis is made in the framework of the QCD improved
parton model~\cite{Altarelli} powered by the DGLAP parton evolution equations.
A part of the analysis was reported in Ref.~\cite{KSUU2009}. Let
\bea
q^i_L(x,Q^2,P^2), \qquad q^\gamma_H(x,Q^2,P^2), \qquad
G^\gamma(x,Q^2,P^2), \qquad \Gamma^{\gamma}(x,Q^2,P^2) ,\label{photonicPDF}
\eea
be light quark (with $i$-flavor and $i=1,\cdots, n_f-1$),
heavy quark, gluon and photon distribution functions in the virtual photon with mass $-P^2$.
Since the parton distributions in the photon are defined in the lowest
order of the electromagnetic coupling constant, $\alpha=e^2/4\pi$,
$\Gamma^{\gamma}$ does not evolve with $Q^2$ and is set to be $\Gamma^{\gamma}(x,Q^2,P^2)=\delta(1-x)$.
In the light quark sector, it is more advantageous to treat, instead of using $q_L^i$, the
``flavor singlet'' and ``nonsinglet'' combinations $q^{\gamma}_{Ls}$ and $q^\gamma_{Lns}$
defined as follows:
\begin{equation}
 q^{\gamma}_{Ls}
\equiv \sum_{i=1}^{n_f - 1} q^{i}_L~, \qquad q^\gamma_{Lns}\equiv
\sum_{i=1}^{n_f - 1} e^2_i \Bigl(q^{i}_L - \frac{1}{n_f - 1}q^{\gamma}_{Ls}\Bigr) , \label{qLs&qiLns}
\end{equation}
where $e_i$ is the electromagnetic charge of the $i$-flavor quark in the unit of proton charge.

Now introducing a row vector
\be
\qv^{\gamma}
= \left( q^{\gamma}_{Ls}, q^{\gamma}_{H},
            G^{\gamma},      q^{\gamma}_{Lns} \right), \label{PartonRowVector}
\ee
the parton distributions $\qv^{\gamma}(x,Q^2,P^2)$ in the virtual photon satisfy the
inhomogeneous DGLAP evolution equation~\cite{Dewitt,GR1983,MVV2002}
\be
\frac{d \qv^{\gamma}(x, Q^2, P^2)}{d\ln Q^2}
= \kv(x,Q^2)+\int_{x}^{1} \frac{dy}{y} \qv^{\gamma}(y, Q^2, P^2)
                  \hat{\Pv}\left(\frac{x}{y}, Q^2 \right) \label{DGLAP}
     ,
\ee
where the elements of a row vector $\kv =
\left( k_{Ls}, k_{H}, k_{G}, k_{Lns} \right)$ refer to the splitting functions of $\gamma$ to the
light flavor-singlet quark combination, to heavy quark, to gluon
and to light flavor-nonsinglet combination, respectively.
The $4 \times 4$ matrix $\hat{\Pv}\left(z, Q^2 \right)$ is expressed as
\begin{eqnarray}
\hat{\Pv} (z, Q^2 )&=&
 \begin{pmatrix}
   P^{S}_{LL}(z, Q^2 ) & P_{HL}(z, Q^2 ) & P_{GL}(z, Q^2 ) & 0 \\
   P_{LH}(z, Q^2 )      & P_{HH}(z, Q^2 ) & P_{GH}(z, Q^2 ) & 0 \\
   P_{LG}(z, Q^2 )      & P_{HG}(z, Q^2 ) & P_{GG}(z, Q^2 ) & 0 \\
   0             & 0        & 0        & P^{NS}_{LL}(z, Q^2 ) \\
 \end{pmatrix},  \label{SplittingMatrix}
\end{eqnarray}
where $P_{AB}$ is a splitting function of $B$-parton to $A$-parton.

The method to solve the above inhomogeneous DGLAP Eq.~(\ref{DGLAP}) is well known~\cite{GR1983}.
We sketch out the procedures.
First we take the Mellin moments,
\begin{eqnarray}
 \frac{d \qv^{\gamma}(n,Q^2,P^2)}{d \ln Q^2}
  &=& \kv(n, Q^2)+\qv^{\gamma}(n, Q^2, P^2)\hat{\Pv}(n, Q^2), \label{MomentDGLAP}
\end{eqnarray}
where we have defined the moments of an arbitrary function $f(x)$ as $f(n)\equiv\int_0^1 dx x^{n-1}f(x)$.
Hereafter we omit the obvious $n$ dependence for simplicity.
Then expansions are made for the splitting functions $\kv(Q^2)$ and $\hat{\Pv}(Q^2)$
in powers of the QCD and QED coupling constants as
\bea
\kv(Q^2)&=&\frac{\alpha}{2\pi}\kv^{(0)}+
\frac{\alpha\alpha_s(Q^2)}{(2\pi)^2}\kv^{(1)} +\cdots , \\
 \hat{\Pv}(Q^2)&=&\frac{\alpha_s(Q^2)}{2\pi} \hat{\Pv}^{(0)}
+\left[\frac{\alpha_s(Q^2)}{2\pi}\right]^2  \hat{\Pv}^{(1)}
+\cdots ,
\eea
and a new variable $t$ is introduced as the evolution variable instead of $Q^2$~\cite{FP1},
\be
t \equiv \frac{2}{\beta_0}\ln\frac{\alpha_s(P^2)}{\alpha_s(Q^2)}~.
\ee

The solution $\qv^\gamma(t)(=\qv^{\gamma}(n,Q^2,P^2))$ of~(\ref{MomentDGLAP}) is
decomposed in the following form:
\be
\qv^\gamma(t)=\qv^{\gamma(0)}(t)+\qv^{\gamma(1)}(t)~,
\ee
where the first and second terms represent the solution in the LO and NLO, respectively.
Then they satisfy the following two differential equations:
\bea
\frac{d {\qv}^{\gamma(0)}(t)}{d t}
 &=& \frac{\alpha}{\alpha_s(t)} {\kv}^{(0)}+{\qv}^{\gamma(0)}(t)P^{(0)}~, \label{APE(0)}\\
\frac{d {\qv}^{\gamma(1)}(t)}{d t}
 &=& \frac{\alpha}{2\pi} \Bigl[ {\bf k}^{(1)}
          -\frac{\beta_1}{2\beta_0} {\bf k}^{(0)}
\Bigr]
+\frac{\alpha_s(t)}{2\pi}{\qv}^{\gamma(0)}(t)
\Bigl[P^{(1)}-\frac{\beta_1}{2\beta_0} P^{(0)} \Bigr]+{\qv}^{\gamma(1)}(t)P^{(0)}~ ,  \label{APE(1)}
\eea
where we have used the fact the QCD effective coupling constant
$\alpha_s(Q^2)$ satisfies
\be
  \frac{d \alpha_s(Q^2)}{d {\rm ln} Q^2}= -\beta_0 \frac{\alpha_s(Q^2)^2}{4\pi}
   -\beta_1 \frac{\alpha_s(Q^2)^3}{(4\pi)^2}+ \cdots~,
\label{beta}
\ee
with $\beta_0=(11-\frac{2}{3}n_f)$ and $\beta_1=(102-\frac{38}{3}n_f)$.
Note that the $P^2$ dependence of $\qv^\gamma$ solely comes from the
initial condition (or boundary condition) as we will see below.

The initial conditions for $\qv^{\gamma(0)}$ and $\qv^{\gamma(1)}$ are obtained as follows:
for $ -p^2= P^2 \gg \Lambda^2$ the photon matrix elements of the hadronic operators $O^n_i$
($i=q_L,H,G,Lns$) can be calculated perturbatively.
(These hadronic operators $O^n_i$ are explained in Sec.~\ref{AnomalousDimension}). Renormalizing at
$\mu^2 = P^2 $, we obtain at one-loop level
\be
  \langle \gamma (p) \mid O^n_i (\mu) \mid \gamma (p) \rangle \vert_{\mu^2= P^2}
= \frac{\alpha}{4\pi}  {\widetilde A}_{n}^{i(1)}~, \qquad i=q_L,H,G,Lns~.
\label{Initial0}
\ee
The ${\widetilde A}_{n}^{i(1)}$ terms represent the operator mixing between the hadronic
operators and photon operators in the NLO and
the operator mixing implies that there exist parton distributions in the photon.
Thus we have, at $\mu^2 = P^2 $ (or at $t=0$),
\be
  {\qv}^{\gamma(0)}(0)=0, \qquad \quad
  {\qv}^{\gamma(1)}(0)= \frac{\alpha}{4\pi} {\widetilde {\Av}}_n^{(1)}~,\label{Initial1}
\ee
with
\be
{\widetilde {\Av}}_n^{(1)}
=\Bigl({\widetilde A}_n^{q_L(1)}, {\widetilde A}_n^{H(1)} ,~ 0,~ {\widetilde A}_n^{Lns(1)}  \Bigr)\label{1loopPME},
\ee
which state that the initial quark distributions emerge not in the LO (the order $\alpha/\alpha_s$)
but in the NLO (the order $\alpha$), and initial gluon distribution starts to emerge
in the NNLO (the order $\alpha\alpha_s$).
Despite of the initial condition ${\qv}^{\gamma(0)}(0)=0$, the LO quark distributions,
both light and heavy, are generated from the photon distribution $\Gamma^\gamma(x,Q^2,P^2)=\delta(1-x)$
(see Eq.~(\ref{photonicPDF})) through the pointlike coupling of the photon to quarks.
The heavy quark parton appears in the LO as a massless quark, while, as we see
in Sec.~\ref{NLOInitialCond}, the heavy quark mass effects
arise from the initial condition ${\qv}^{\gamma(1)}(0)$
(more closely, ${\widetilde A}_n^{H(1)}$ in Eq.~\eqref{1loopPME}) in the NLO.

With these initial conditions~(\ref{Initial1}), the solutions
${\qv}^{\gamma(0)}(t)$ and ${\qv}^{\gamma(1)}(t)$ are given by
\bea
{\qv}^{\gamma(0)}(t)&=&\frac{4\pi}{\alpha_s(t)}~ {\bf a}~
     \Biggl\{1-\biggl[\frac{\alpha_s(t)}{\alpha_s(0)}
   \biggr]^{1-\frac{2 P^{(0)}}{\beta_0}}  \Biggr\} , \label{SolLO} \\
 \nonumber  \\
{\qv}^{\gamma(1)}(t)&=& -2~{\bf a}~
  \Biggl\{  \int^t_0 d\tau  e^{( P^{(0)}-\frac{\beta_0}{2})\tau}~
\Bigl[ P^{(1)}-\frac{\beta_1}{2\beta_0} P^{(0)} \Bigr]~
              e^{- P^{(0)}\tau }
  \Biggr\}~ e^{ P^{(0)}t}   \nonumber  \\
& &+ {\bf b}~\Biggl\{1-\biggl[\frac{\alpha_s(t)}{\alpha_s(0)}
   \biggr]^{-\frac{2 P^{(0)}}{\beta_0}} \Biggr\}
                +{\qv}^{\gamma(1)}(0)\biggl[\frac{\alpha_s(t)}{\alpha_s(0)}
   \biggr]^{-\frac{2 P^{(0)}}{\beta_0}} , \label{SolNLO}
\eea
where
\bea
  {\bf a}&=&\frac{\alpha}{2\pi\beta_0} {\bf k}^{(0)}
\frac{1}{1-\frac{2 P^{(0)}}{\beta_0}} , \\
{\bf b}&= &\Biggl\{ \frac{\alpha}{2\pi} \Bigl[ {\bf k}^{(1)}
          -\frac{\beta_1}{2\beta_0} {\bf k}^{(0)}
\Bigr] +2~{\bf a}~\Bigl[P^{(1)}-\frac{\beta_1}{2\beta_0} P^{(0)} \Bigr]
\Biggr\}\frac{-1}{ P^{(0)}} .
\eea

The moments of the splitting functions are related to the
anomalous dimensions of operators as follows:
\bea
    P^{(0)}&=&-\frac{1}{4}\widehat \gamma^{(0)}_n~, \qquad \qquad
    P^{(1)}=-\frac{1}{8}\widehat \gamma^{(1)}_n~,  \\
   {\kv}^{(0)}&=&\frac{1}{4}{\Kv}^{(0)}_n, \qquad \qquad \ \
    {\kv}^{(1)}=\frac{1}{8}{\Kv}^{(1)}_n~,
\eea
where $\widehat \gamma^{(0)}_n$ and $\widehat \gamma^{(1)}_n$ are the $4 \times 4$ one-loop
and two-loop anomalous dimension matrices in the hadronic sector, respectively, and
${\Kv}^{(0)}_n$ and ${\Kv}^{(1)}_n$ are the four-component row vectors which represent
the mixing in one-loop and two-loop level, respectively, between the photon operator and
the four hadronic operators. The details are explained in the next section.

The evaluation of ${\qv}^{\gamma(0)}(t)$ and ${\qv}^{\gamma(1)}(t)$
in Eqs.~(\ref{SolLO}) and~(\ref{SolNLO}) can be easily done
by introducing the projection operators $P^n_i$ such as
\bea
     P^{(0)}&=&-\frac{1}{4}\widehat \gamma^0_n =-\frac{1}{4}\sum_{i=\psi,+,-,Lns}
\lambda^n_i~P^n_i,     \qquad   i=\psi,+,-,Lns,  \\
   P^n_i~P^n_j&=&\begin{cases}
             0 &i\ne j, \\
                        P^n_i & i=j
     \end{cases},    \qquad
  \sum_{i=\psi,+,-,Lns}  P^n_i ={\bf 1} ,
\eea
where $\lambda^n_i$ are the four eigenvalues of the matrix $\widehat \gamma^0_n$.
Then, rewriting
$\alpha_s(0)$ and $\alpha_s(t)$ as $\alpha_s(P^2)$ and $\alpha_s(Q^2)$, respectively, we obtain
\bea
  {\qv}^{\gamma(0)}(t)/\Bigl[ \frac{\alpha}{8\pi \beta_0}\Bigr] &=&
   \frac{4\pi}{\alpha_s(Q^2)}~{\Kv}^{(0)}_n~\sum_i P^n_i~
       \frac{1}{1+d_i^n}
   \Biggl\{1-\biggl[\frac{\alpha_s(Q^2)}{\alpha_s(P^2)}
   \biggr]^{1+d_i^n}  \Biggr\}~, \label{qgammaZERO}  \\
&&\nn\\
 {\qv}^{\gamma(1)}(t)/\Bigl[ \frac{\alpha}{8\pi \beta_0}\Bigr] &=&\Biggl\{{\Kv}^{(1)}_n~\sum_i P^n_i~
\frac{1}{d_i^n} + \frac{\beta_1}{\beta_0}{\Kv}^{(0)}_n
~\sum_i P^n_i~\Bigl( 1 - \frac{1}{d_i^n} \Bigr)  \nonumber  \\
&- & {\Kv}^{(0)}_n \sum_{j,i} \frac{P^n_j~ \widehat \gamma^{(1)}_n~P^n_i}
  {2\beta_0+\lambda^n_j-\lambda^n_i}~\frac{1}{d_i^n}~
   - 2\beta_0{\widetilde {\Av}}_n^{(1)}~\sum_i P^n_i  \Biggr\} \nonumber  \\
& & \qquad \qquad \qquad \qquad \qquad \qquad \qquad \times
\Biggl\{1-\biggl[\frac{\alpha_s(Q^2)}{\alpha_s(P^2)}
   \biggr]^{d_i^n}  \Biggr\}\nonumber  \\
&+&  \Biggl\{{\Kv}^{(0)}_n\sum_{i,j}~\frac{ P^n_i~\widehat \gamma^{(1)}_n
~P^n_j}{2\beta_0+\lambda^n_i-\lambda_j^n}~
\frac{1}{1+d_i^n}
 - \frac{\beta_1}{\beta_0}{\Kv}^{(0)}_n~
\sum_i~P^n_i~\frac{d_i^n}{1+d_i^n}
\Biggr\} \nonumber  \\
& & \qquad \qquad \qquad \qquad \qquad \qquad \qquad \times
\Biggl\{1-\biggl[\frac{\alpha_s(Q^2)}{\alpha_s(P^2)}
   \biggr]^{1+d_i^n}  \Biggr\} \nonumber  \\
  &+& 2\beta_0{\widetilde {\Av}}_n^{(1)},\label{qgammaONE}
\eea
where $d_i^n\equiv \frac{\lambda_i^n}{2\beta_0}$~ and $i, j=\psi,+,-,Lns$.

Finally, since
$\qv^\gamma(t)=\qv^{\gamma(0)}(t)+\qv^{\gamma(1)}(t)$, and from~(\ref{PartonRowVector}),
the moments for the parton distributions of
the ``flavor-singlet'' light quark, heavy quark, gluon and ``flavor-nonsinglet'' light quark
are given, respectively, by
\bea
 q_{Ls}^\gamma(n,Q^2,P^2)=(1,1)\ {\rm component\ of\ the\  row\ vector\ }\qv^\gamma(t)~, \label{Mofqls}\\
 q_H^\gamma(n,Q^2,P^2)=(1,2)\ {\rm component\ of\ the\  row\ vector\ }\qv^\gamma(t)~, \label{MofH}\\
 G^\gamma(n,Q^2,P^2)=(1,3)\ {\rm component\ of\ the\  row\ vector\ }\qv^\gamma(t)~, \label{MofG}\\
 q_{Lns}^\gamma(n,Q^2,P^2)=(1,4)\ {\rm component\ of\ the\  row\ vector\ }\qv^\gamma(t)~.\label{Mofqlns}
\eea

\bigskip

\section{Parameters in ${\qv}^{\gamma(0)}(t)$ and ${\qv}^{\gamma(1)}(t)$}
\label{Parameters}

We give here the information on the parameters which appear in ${\qv}^{\gamma(0)}(t)$ and
${\qv}^{\gamma(1)}(t)$ in~(\ref{qgammaZERO})-(\ref{qgammaONE}). They are calculated in
 $\rm{\overline{MS}}$ scheme~\cite{BBDM}. We introduce the following quark-charge factors
in the massless quark sector:
\be
\langle e^2 \rangle_{L}\equiv\frac{1}{n_f-1}\sum_{i=1}^{n_f-1}e_i^2~,\qquad
\langle e^4 \rangle_{L}\equiv\frac{1}{n_f-1}\sum_{i=1}^{n_f-1}e_i^4~. \label{ChargeFactor}
\ee

\subsection{Anomalous dimensions}
\label{AnomalousDimension}

Corresponding to the splitting functions given in Eq.~(\ref{SplittingMatrix}),
the anomalous dimensions in the hadronic sector are expressed in the form of
$4\times 4$ matrix as
\be
{\widehat \gamma}_{n}(g)=
\left(
\begin{array}{cccc}
\gamma_{q_Lq_L}^{n}(g)\ &\gamma_{H q_L}^{n}(g)\ &\gamma_{G q_L}^{n}(g)&0\\
\gamma_{q_L H}^{n}(g)\ &\gamma_{HH}^{n}(g)\ &\gamma_{GH}^{n}(g)&0\\
\gamma_{q_L G}^{n}(g)&\gamma_{H G}^{n}(g)&
\gamma_{GG}^{n}(g)&0\\
0&0&0&\gamma_{Lns}^{n}(g)
\end{array}
\right)~.
\ee
The four-component row vector
\be
   {\Kv}_{n}(g,\alpha)=\left(K^n_{q_L}(g,\alpha), \ K^n_{H}(g,\alpha), \
    K^n_{G}(g,\alpha), \ K^n_{Lns}(g,\alpha)\right) ,
\ee
represents the mixing between photon and four hadronic operators.
Here an importance of inclusion of the heavy quark operator should be stressed.
We treat the heavy quark in the same way as the light quarks and assume that
both heavy and light quarks are radiatively generated from the photon target.
In contrast, in the case of the nucleon target, heavy quarks are treated as radiatively generated
from the gluon and light quarks.

Since the elements $\gamma_{q_L H}^{n}$ and $\gamma_{H q_L}^{n}$
start at the order of $\alpha_s^2$ and $\gamma_{q_L G}^{n}$ has a factor $(n_f-1)$,
the one-loop anomalous dimension matrix ${\widehat \gamma}^{(0)}_{n}$ is
expressed as
\be
{\widehat \gamma}^{(0)}_{n}=
\left(
\begin{array}{ccc|c}
\gamma_{\psi\psi}^{(0),n}\ &0\ &\gamma_{G\psi}^{(0),n}&0\\
0\ &\gamma_{\psi\psi}^{(0),n}\ &\gamma_{G\psi}^{(0),n}&0\\
\frac{n_f-1}{n_f}\gamma_{\psi G}^{(0),n}&\frac{1}{n_f}\gamma_{\psi G}^{(0),n}&
\gamma_{GG}^{(0),n}&0\\
\hline
0&0&0&\gamma_{\psi\psi}^{(0),n}
\end{array}
\right)~,
\ee
where $\gamma_{\psi\psi}^{(0),n}$, $\gamma_{\psi G}^{(0),n}$, $\gamma_{G\psi}^{(0),n}$ and
$\gamma_{GG}^{(0),n}$ are well-known one-loop anomalous dimensions for the hadronic sector
which appear when all $n_f$ flavor quarks are massless, and are given, for example, in
Eqs.~(4.1), (4.2), (4.3) and~(4.4) of Ref.~\cite{BB}.
The four eigenvalues of ${\widehat \gamma}^{(0)}_{n}$ are
\begin{eqnarray}
\lambda_\psi^n&=&\gamma_{\psi\psi}^{(0),n},\\
\lambda_\pm^n&=&\frac{1}{2}\left\{
\gamma_{\psi\psi}^{(0),n}+\gamma_{GG}^{(0),n}\pm
\sqrt{(\gamma_{\psi\psi}^{(0),n}-\gamma_{GG}^{(0),n})^2+4\gamma_{\psi G}^{(0),n}
\gamma_{G\psi}^{(0),n}}\right\}~,\\
\lambda_{Lns}^n&=&\gamma_{\psi\psi}^{(0),n}~.
\end{eqnarray}
The one-loop anomalous dimension matrix ${\widehat \gamma}^{(0)}_{n}$ can be
expressed in terms of its eigenvalues $\lambda^n_i~ (i=\psi,+, -, Lns)$ and
corresponding projection operators as
\be
  {\widehat \gamma}^{(0)}_{n}=\sum_{i=\psi,+, -, Lns} \lambda^n_i~P^n_i ,
\ee
with
\begin{eqnarray}
&&\hspace{-0.5cm}P_\psi^n=
\left(
\begin{array}{ccc|c}
\frac{1}{n_f}\ &-\frac{1}{n_f}\ &0&0\\
-\frac{n_f-1}{n_f}\ &\frac{n_f-1}{n_f}\ &0&0\\
0&0&0&0\\
\hline
0&0&0&0
\end{array}
\right)~,\\
&&\hspace{-0.5cm}P_\pm^n=
\frac{1}{\lambda_\pm^n-\lambda_\mp^n}
\left(
\begin{array}{ccc|c}
\frac{n_f-1}{n_f}(\gamma_{\psi\psi}^{(0),n}-\lambda_\mp^n) &
\frac{1}{n_f}(\gamma_{\psi\psi}^{(0),n}-\lambda_\mp^n)\ &\gamma_{G\psi}^{(0),n}&0\\
\frac{n_f-1}{n_f}(\gamma_{\psi\psi}^{(0),n}-\lambda_\mp^n) &
\frac{1}{n_f}(\gamma_{\psi\psi}^{(0),n}-\lambda_\mp^n)\ &\gamma_{G\psi}^{(0),n}&0\\
\frac{n_f-1}{n_f}\gamma_{\psi G}^{(0),n}\ &\frac{1}{n_f}\gamma_{\psi G}^{(0),n}
&\gamma_{GG}^{(0),n}-\lambda_\mp^n&0\\
\hline
0&0&0&0
\end{array}
\right)~,\\
&&\hspace{-0.5cm}P_{Lns}^n=
\left(
\begin{array}{ccc|c}
0\  &0\ &0\ &0 \\
0\  &0\ &0\ &0 \\
0\  &0\ &0\ &0 \\
\hline
0&0&0&1
\end{array}
\right)~.
\end{eqnarray}

The elements of the one-loop anomalous dimension row vector
$\Kv_{n}^{(0)}=\left( K_{q_L}^{(0),n},  K_{H}^{(0),n},  K_{G}^{(0),n},  K_{Lns}^{(0),n}
     \right)$ are given by
\begin{eqnarray}
 K_{q_L}^{(0),n} &=& 24 (n_f\!-\!1) \langle e^2 \rangle_{L}~ k_{n}^{0},\\
%---
 K_{H}^{(0),n} &=& 24 e_{H}^2~ k_{n}^{0},\\
%---
 K_{G}^{(0),n} &=& 0,\\
%---
 K_{Lns}^{(0),n} &=& 24 (n_f\!-\!1)
  \left(   \langle e^4 \rangle_{L} - (\langle e^2 \rangle_{L})^2 \right) ~k_{n}^{0},
\end{eqnarray}
with
\be
  k_{n}^{0} = \frac{n^2+n+2}{n(n+1)(n+2)}. \label{k0n}
\ee

The two-loop anomalous dimensions for hadronic sector with a heavy quark are inferred
from those for the case when all $n_f$-flavor quarks are massless~\cite{FRS}.
Minor changes of group factors arise from quark loops:
\bea
\gamma^{(1),n}_{Lns}&=&\gamma^{(1),n}_{NS}~,\\
\gamma^{(1),n}_{q_Lq_L}&=&\gamma^{(1),n}_{NS}+C_F \Bigl(\frac{n_f-1}{2}\Bigr)
 D^{n}_{PS,\psi\psi}~,\\
\gamma^{(1),n}_{q_L H}&=&C_F \Bigl(\frac{n_f-1}{2}\Bigr)
 D^{n}_{PS,\psi\psi}~,\\
\gamma^{(1),n}_{q_L G}&=&8 C_F \Bigl(\frac{n_f-1}{2}\Bigr)~ D^n_{\psi G}+8 C_A \Bigl(\frac{n_f-1}{2}\Bigr)~ E^n_{\psi G}~,\\
\gamma^{(1),n}_{H q_L}&=&C_F \Bigl(\frac{1}{2}\Bigr)
 D^{n}_{PS,\psi\psi}~,\\
\gamma^{(1),n}_{H H}&=&\gamma^{(1),n}_{NS}+C_F \Bigl(\frac{1}{2}\Bigr)
 D^{n}_{PS,\psi\psi}~,\\
\gamma^{(1),n}_{H G}&=&8 C_F \Bigl(\frac{1}{2}\Bigr)~ D^n_{\psi G}+8 C_A \Bigl(\frac{1}{2}\Bigr)~ E^n_{\psi G}~,
\\
\gamma^{(1),n}_{G q_L}&=&\gamma^{(1),n}_{G \psi}~,\\
\gamma^{(1),n}_{G H}&=&\gamma^{(1),n}_{G \psi}~,
\eea
where $C_A=3$ and $C_F=\frac{4}{3}$ in QCD. The
anomalous dimensions $\gamma^{(1),n}_{NS}$ and $\gamma^{(1),n}_{G \psi}$
for the case of $n_f$ massless quarks are given, for example,
in Eq.~(3.5) of Ref.~\cite{MVV2004NNLOpart1} and Eq.~(3.8) of Ref.~%
\cite{MVV2004NNLOpart2}, respectively (see also Ref.~\cite{USU2007}).
The expression of the ``pure singlet'' contribution $ D^{n}_{PS,\psi\psi}$
and those of $ D^n_{\psi G}$ and $ E_{\psi G}^n$
(which appear in the two-loop anomalous dimension $\gamma^{(1),n}_{\psi G}$
in the case of $n_f$ massless quarks) are
given in Eqs.~(3.6) and~(3.7) of Ref.~\cite{MVV2004NNLOpart2}, respectively.
The anomalous dimension $\gamma^{(1),n}_{G G}$ for the case with a heavy quark
is the same with the case of $n_f$ massless quarks and is given in (3.9) of Ref.~\cite{MVV2004NNLOpart2}.

The elements of the two-loop anomalous dimension row vector
$\Kv_{n}^{(1)}
 = \left( K_{q_L}^{(1),n},  K_{H}^{(1),n},
            K_{G}^{(1),n},     K_{Lns}^{(1),n}
     \right)$
are given by
\bea
 K_{q_L}^{(1),n}
  &=& -3 (n_f-1) \langle e^2 \rangle_{L}C_F~8D^n_{\psi G},\\
%---
 K_{H}^{(1),n}
  &=& -3~ e_{H}^2~C_F~8D^n_{\psi G},\\
%---
 K_{G}^{(1),n} &=& -3~\Bigl((n_f-1)\langle e^2 \rangle_{L}+e_H^2\Bigr)C_F~8(D_{GG}^n-1),
      \label{KG(1)}\\
%---
 K_{Lns}^{(1),n}
  &=& -3 (n_f-1)
  \left(   \langle e^4 \rangle_{L}
         - (\langle e^2 \rangle_{L})^2 \right) ~C_F~8D^n_{\psi G},
\eea
where $D_{GG}^n$ is obtained from $\gamma^{(1),n}_{G G}$ by replacing
$C_A \rightarrow 0$ and $C_Fn_f \rightarrow \frac{1}{4}$.

\subsection{The one-loop photon matrix elements}
\label{NLOInitialCond}

\begin{figure}[htb]
  \begin{center}
    \includegraphics[scale=0.6]{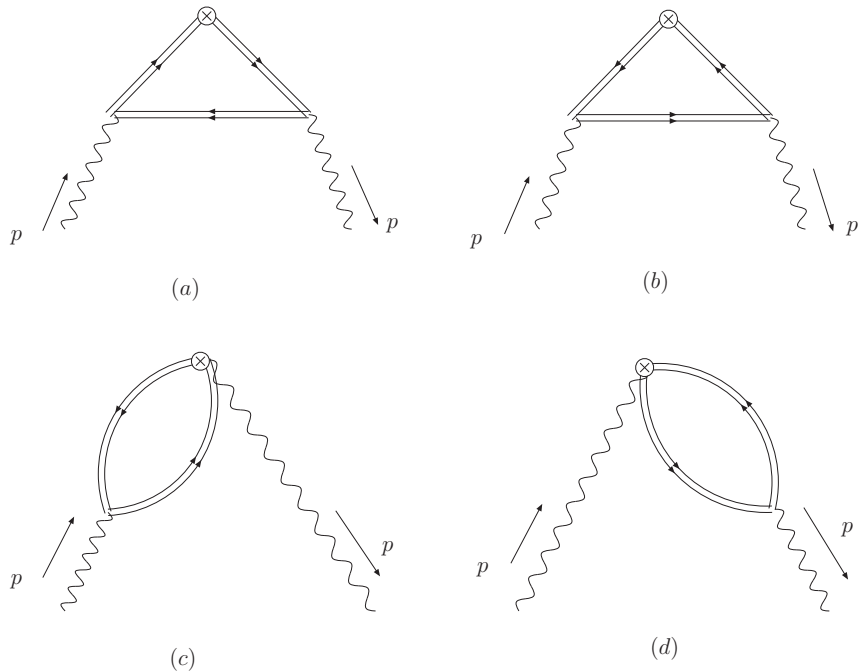}
    \caption{The diagrams for ${\widetilde A}_n^{H(1)}$.
             The double lines express the heavy quark.}
    \label{fig_OME}
  \end{center}
\end{figure}

The elements of the row vector ${\widetilde {\Av}}_n^{(1)}$ in Eq.~(\ref{1loopPME})
are given by
\bea
{\widetilde A}_n^{q_L(1)}&=&3(n_f\!-\!1)\langle e^2 \rangle_{L}H_q^{(1)}(n),\\
{\widetilde A}_n^{H(1)}&\equiv&3~e_H^2H_q^{(1)}(n)+\Delta {\widetilde A}^n_H
\label{DefDeltaAH},\\
{\widetilde A}_n^{Lns(1)}&=&3(n_f\!-\!1)\Bigl(\langle e^4 \rangle_{L}-(\langle e^2 \rangle_{L})^2\Bigr)H_q^{(1)}(n),
\eea
where
\bea
H_q^{(1)}(n)&=&4 \biggl[  \frac{n^2+n+2}{n(n+1)(n+2)} S_1(n) +
\frac{4}{(n+2)^2}-\frac{4}{(n+1)^2}+\frac{1}{n^2}-\frac{1}{n}  \biggr]~,\\
 \Delta \tilde{A}_{H}^{n}
 &=& 3 e_{H}^2 \cdot
     4 \left[ - \frac{n^2+n+2}{n(n+1)(n+2)}
               \ln \frac{m^2}{P^2}
	     + \frac{1}{n} - \frac{1}{n^2}
	     + \frac{4}{(n+1)^2} - \frac{4}{(n+2)^2}
	     - \frac{n^2+n+2}{n(n+1)(n+2)} S_1(n)
       \right]~,\label{deltaAH}
\eea
with $S_1(n)=\sum_{j=1}^n \frac{1}{j}$, and ${\widetilde A}_n^{H(1)}$ is obtained by
evaluating the diagrams in Fig.~\ref{fig_OME} in the limit $\Lambda^2 \ll P^2 \ll m^2$.
The heavy quark mass effects reside in the term $\Delta \tilde{A}_{H}^{n}$.

%%%%%%%%%%%%%%%%%%%%%%%%%%%%%%%%%%

\section{The NLO PDFs in DIS$_\gamma$ Scheme}
\label{DISg-scheme}

The structure functions of the photon (nucleon) are expressed as convolutions of coefficient functions
and parton distributions of the target photon (nucleon).
But it is well known that these coefficient functions and parton distributions are
by themselves factorization-scheme dependent.
The relevant quantities given in Sec.~\ref{Parameters} were calculated
in $\rm{\overline{MS}}$ scheme. When we insert them into the formulae
given by~(\ref{qgammaZERO}) and~(\ref{qgammaONE}), we obtain the parton
distributions predicted by $\rm{\overline{MS}}$ scheme.
Meanwhile, an interesting and also useful factorization scheme called DIS$_\gamma$ was introduced in the
analysis of the photon structure function $F_2^\gamma$~\cite{GRV1992a}. In this scheme, the
photonic coefficient function $C_2^\gamma$, i.e., the direct photon contribution to
$F_2^\gamma$, is absorbed into the photonic quark
distributions.

The Mellin moments of the virtual photon structure function $\frac{1}{x}F_2^{\gamma}(x,Q^2,P^2)$ is expressed as
\bea
     \int_0^1 dx x^{n-1}\frac{1}{x}~  F_2^{\gamma}(x,Q^2,P^2)
&\equiv&F_2^{\gamma}(n,Q^2,P^2)  \nonumber \\
&=& \bm{q}^{\gamma}(n,Q^2,P^2)\cdot \bm{C}_2(n,Q^2) +C_2^\gamma(n,Q^2)
\label{F2moment}~,
\eea
where $\bm{q}^\gamma(n,Q^2,P^2)$ is the four-component row vector given in~(\ref{PartonRowVector}). The column vector
$\bm{C}_2(n,Q^2)$ is made up of four hadronic coefficient functions,
\bea
   \bm{C}_2(n,Q^2)&\equiv&( C_2^{Ls}(n,Q^2),~C_2^{H}(n,Q^2),~
  C_2^G(n,Q^2),~C_2^{Lns}(n,Q^2)
   )^{\rm T} , \label{Coefficient}
\eea
where $C_2^{Ls}, C_2^{H}, C_2^G$ and $C_2^{Lns}$
are coefficient functions corresponding to the light ``flavor-singlet'' quark,
heavy quark, gluon and light ``flavor-nonsinglet'' quark, respectively. The last term
$C_2^\gamma(n,Q^2)$ in~(\ref{F2moment}) is the photonic coefficient function.
The moments of the parton distributions in ${\rm DIS}_{\gamma}$ scheme
are obtained as follows~\cite{GRV1992a,MVV2002}. In this scheme, the hadronic coefficient functions
are the same as their counterparts in ${\overline {\rm MS}}$ scheme,
but the photonic coefficient function is absorbed into the quark
distributions and thus set to zero,
\begin{equation}
{\bm{C}}_{2}(n,Q^2)|_{{\rm DIS}_{\gamma}}={\bm{C}}_{2}(n,Q^2)|_{\overline {\rm MS}}~,
\qquad C_{2}^{\gamma}(n,Q^2)|_{{\rm DIS}_{\gamma}}=0.
\end{equation}
Then Eq.~(\ref{F2moment}) gives
\bea
F_2^{\gamma}(n,Q^2,P^2)
&=&\bm{q}^{\gamma}(n,Q^2,P^2)|_{{\rm DIS}_{\gamma}} \cdot \bm{C}_2(n,Q^2)|_{{\rm DIS}_{\gamma}}
\nonumber\\
&=&\bm{q}^{\gamma}(n,Q^2,P^2)|_{{\rm DIS}_{\gamma}}\cdot
\bm{C}_2(n,Q^2)|_{\overline{\rm MS}}~. \label{F2DISgamma}
\eea
On the other hand, $F_2^{\gamma}(n,Q^2,P^2)$ is expressed in ${\overline {\rm MS}}$ scheme as
\begin{equation}
F_2^{\gamma}(n,Q^2,P^2) =\bm{q}^{\gamma}(n,Q^2,P^2)|_{\overline{\rm MS}}\cdot
\bm{C}_2(n,Q^2)|_{\overline{\rm MS}}+ C_2^\gamma(n,Q^2)|_{\overline{\rm MS}}~.\label{F2MSbar}
\end{equation}

We expand $\qv^{\gamma}(n,Q^2,P^2)|_{{\rm DIS}_{\gamma}}$
in terms of the LO and NLO distributions and
$\Cv_2(n,Q^2)|_{\overline{\rm MS}}$ in powers of $\alpha_s(Q^2)$ up to the NLO
as follows:
\bea
\qv^{\gamma}(n,Q^2,P^2)|_{{\rm DIS}_{\gamma}}&=&\qv_n^{\gamma(0)}
+\qv_n^{\gamma(1)}|_{{\rm DIS}_{\gamma}}+\cdots~,
\label{PartonExpanded}\\
\Cv_2(n,Q^2)|_{\overline{\rm MS}}&=&\Cv_{2,n}^{(0)}
+\frac{\alpha_s(Q^2)}{4\pi}\Cv_{2,n}^{(1)}|_{\overline{\rm MS}} +\cdots~, \label{CoefficientExpanded}
\eea
where the LO $\qv_n^{\gamma(0)}$ and $\Cv_{2,n}^{(0)}$ are both factorization-scheme independent.
Denoting the difference of $\qv_n^{\gamma(1)}|_{{\rm DIS}_{\gamma}}$
from ${\overline {\rm MS}}$ scheme prediction as $\delta{\qv}_n^{\gamma(1)}|_{{\rm DIS}_{\gamma}}$, we write
\begin{equation}
{\qv}_n^{\gamma(1)}|_{{\rm DIS}_{\gamma}}\equiv
{\qv}_n^{\gamma(1)}|_{\overline {\rm MS}}+\delta{\qv}_n^{\gamma(1)}|_{{\rm DIS}_{\gamma}} ~.
\label{SchemeDifference}
\end{equation}
Putting~(\ref{PartonExpanded})-(\ref{SchemeDifference})
into the r.h.s. of~(\ref{F2DISgamma}) and comparing the result
with~(\ref{F2MSbar}), we find
\begin{equation}
C_2^\gamma(n,Q^2)|_{\overline{\rm MS}}
=\delta\qv_n^{\gamma(1)}|_{{\rm DIS}_{\gamma}}\cdot \Cv_{2,n}^{(0)}+\cdots~. \label{CoefficientGammaExpanded}
\end{equation}
Since
\be
\Cv_{2,n}^{(0)}=\left(\langle e^2 \rangle_L, e_H^2, 0, 1 \right)^{\rm T}~,
\ee
the r.h.s. of Eq.~(\ref{CoefficientGammaExpanded}) is rewritten as
\bea
\delta\qv_n^{\gamma(1)}|_{{\rm DIS}_{\gamma}}\cdot \Cv_{2,n}^{(0)}
&=&\langle e^2 \rangle_L~ \delta q_{Ls,n}^{\gamma(1)}|_{{\rm DIS}_{\gamma}}
+e_H^2 \delta q_{H,n}^{\gamma(1)}|_{{\rm DIS}_{\gamma}} +
\delta q_{Lns,n}^{\gamma(1)}|_{{\rm DIS}_{\gamma}}~. \label{RHSCoefficientGammaExpanded}
\eea
The moment of the photonic coefficient function $C_2^\gamma(n,Q^2)|_{\overline{\rm MS}}$
is written up to the one-loop level as
\begin{equation}
C_2^\gamma(n,Q^2)|_{\overline{\rm MS}}
=\frac{\alpha}{4\pi}3 \Bigl\{(n_f-1)\langle e^4 \rangle_L B_{\gamma}^{L,n} +e_H^4 B_{\gamma}^{H,n}\Bigr\} +\cdots ~.
\label{CoefficientGammaNLO}
\end{equation}
Now dividing the light quark-charge factor $\langle e^4 \rangle_L$ into two parts,
the light flavor-singlet and nonsinglet parts, as
\begin{equation}
\langle e^4 \rangle_L
=\langle e^2 \rangle_L\langle e^2 \rangle_L +\left(\langle e^4 \rangle_L-(\langle e^2 \rangle_L)^2 \right)~,
\end{equation}
and from Eqs.~(\ref{RHSCoefficientGammaExpanded}) and~(\ref{CoefficientGammaNLO}) we find at the NLO
\bea
\delta q^{\gamma (1)}_{Ls,n}|_{{\rm DIS}_{\gamma}}&=&\frac{\alpha}{4\pi}3
(n_f-1)\langle e^2 \rangle_L~
B_{\gamma}^{L,n}~, \label{deltaqLs(1)}\\
\delta q^{\gamma (1)}_{H,n}|_{{\rm DIS}_{\gamma}}&=& \frac{\alpha}{4\pi} 3~ e_H^2
~B_{\gamma}^{H,n}~, \\
\delta q^{\gamma (1)}_{Lns,n}|_{{\rm DIS}_{\gamma}}&=&\frac{\alpha}{4\pi}
3(n_f-1)\left(\langle e^4 \rangle_L-(\langle e^2 \rangle_L)^2 \right)
B_{\gamma}^{L, n}~~. \label{deltaqLns(1)}
\eea
The coefficient $B_{\gamma}^{L,n}$ is related to the one-loop
gluon coefficient ${\overline B}_G^n$ by
$B_{\gamma}^{L,n}=\frac{2}{n_f}{\overline B}_G^n$~\cite{BB}, and
given by
\be
B_{\gamma}^{L,n} =
4\left[  - \frac{n^2+n+2}{n(n+1)(n+2)} \left(1 + S_{1}(n) \right)
                  + \frac{4}{(n+1)} - \frac{4}{(n+2)}
                  + \frac{1}{n^2}
          \right],
\ee
while $B_\gamma^{H,n}$ is calculated in the heavy quark mass limit ($\Lambda^2 \ll P^2 \ll m^2$) and we find
\be
B_\gamma^{H,n}=B_\gamma^{L,n}~.
\ee

Finally in ${\rm DIS}_{\gamma}$ scheme we set in all orders
\begin{equation}
G^{\gamma}(n,Q^2,P^2)|_{{\rm DIS}_{\gamma}}=G^{\gamma}(n,Q^2,P^2)|_{\overline{\rm MS}}~.
\label{GDISg}
\end{equation}

%%%%%%%%%%%%%%%%%%%%%%%%%%%%%%%%%%%%%%%%%%%%
\section{Numerical analysis for PDFs with heavy quark effects}
\label{result}

The parton distributions are recovered from their moments
by the inverse Mellin transformation.
Using the formulae given in Eqs.~\eqref{qgammaZERO}-\eqref{Mofqlns}
and parameters enumerated in Sec.~\ref{Parameters},
we obtain the parton distribution functions in the virtual photon
in \MSbar{} scheme up to the NLO .
We have considered the following two cases:
\begin{enumerate}
  \item[(i)]  $Q^2= 5\GeV^2$ and $P^2=0.35\GeV^2$,
  \item[(ii)] $Q^2=30\GeV^2$ and $P^2=0.35\GeV^2$.
\end{enumerate}
In both cases we take $n_f=4$, and choose $c$ as a heavy quark
and assume that the other $u$, $d$ and $s$-quarks are massless.
We take $m_c=1.3\GeV$ as an input for the charm quark mass and
 put $\Lambda=0.2\GeV$ for the QCD scale parameter.

\begin{figure}
  \begin{center}
    \def\SCALE{0.65}
    \def\OFFSET{27pt}
    \begin{tabular}{cc}
      \includegraphics[scale=\SCALE]{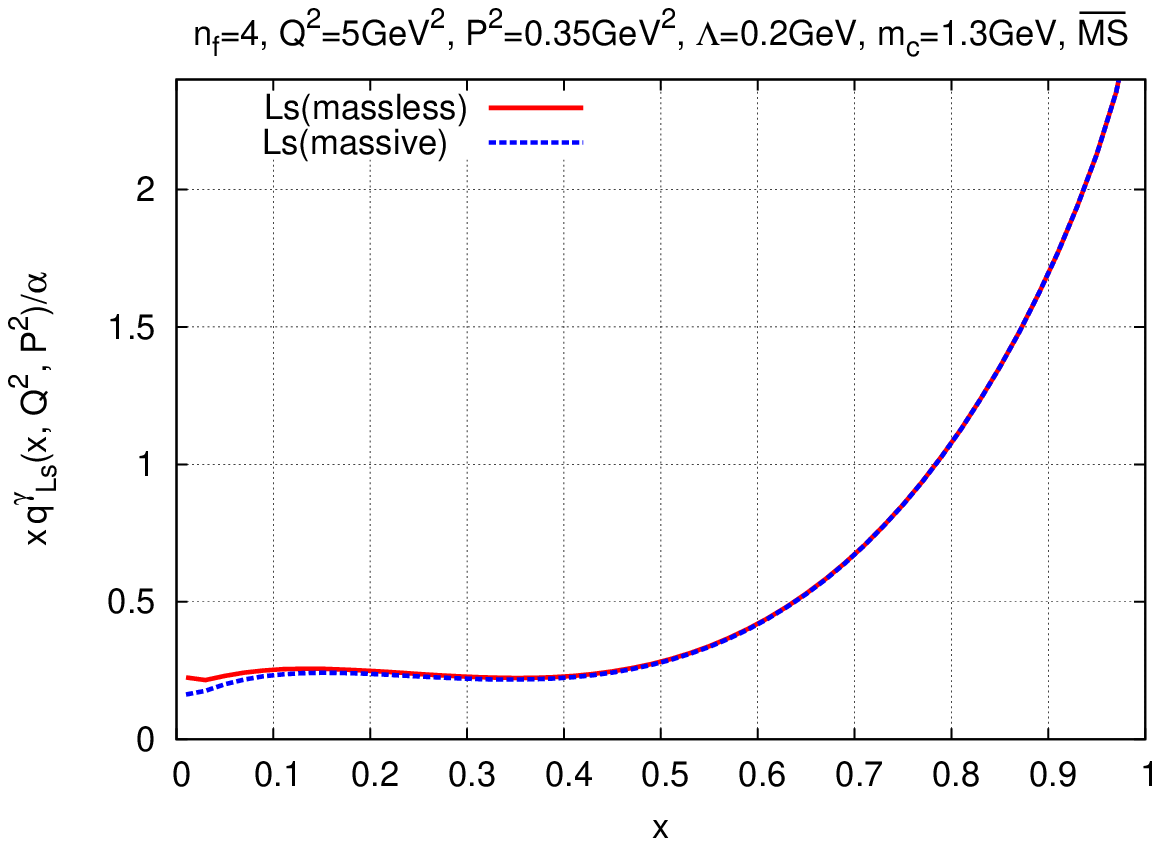} &
      \includegraphics[scale=\SCALE]{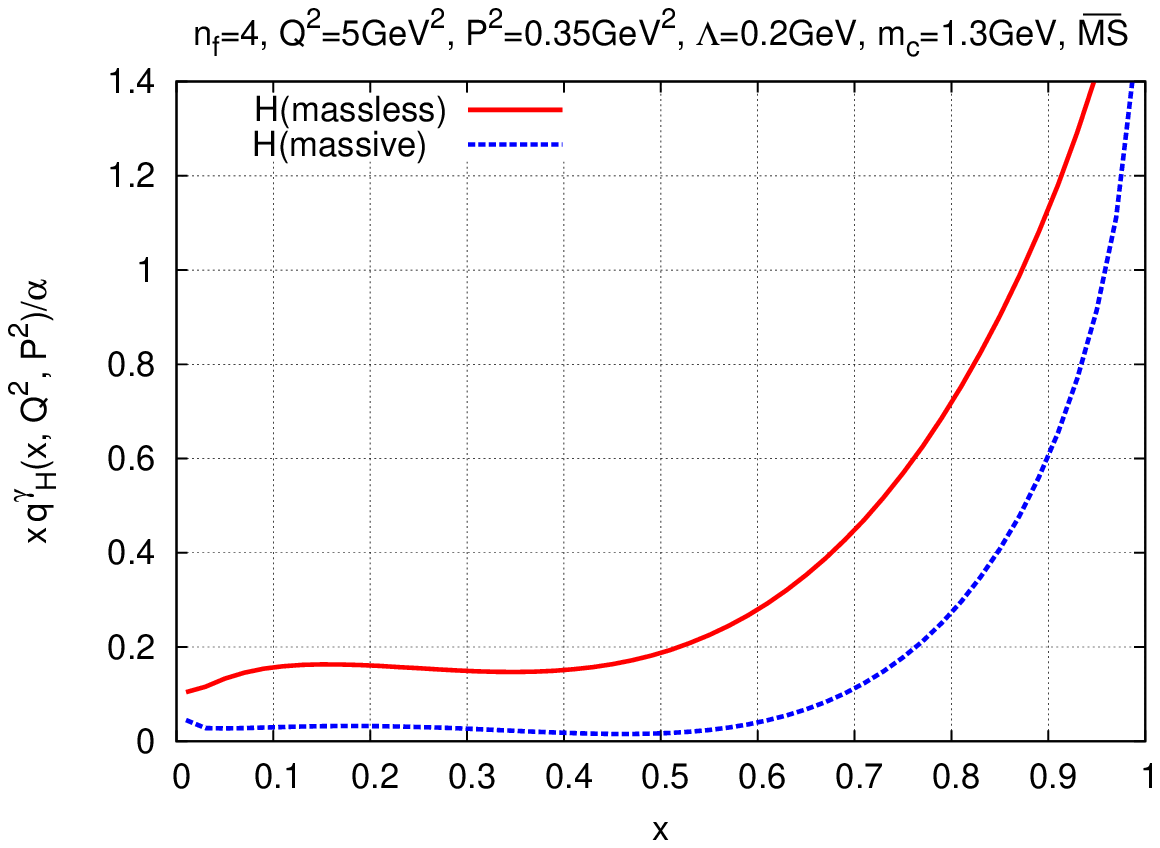} \\
      \hspace{\OFFSET} (a) & \hspace{\OFFSET} (b) \\
      \includegraphics[scale=\SCALE]{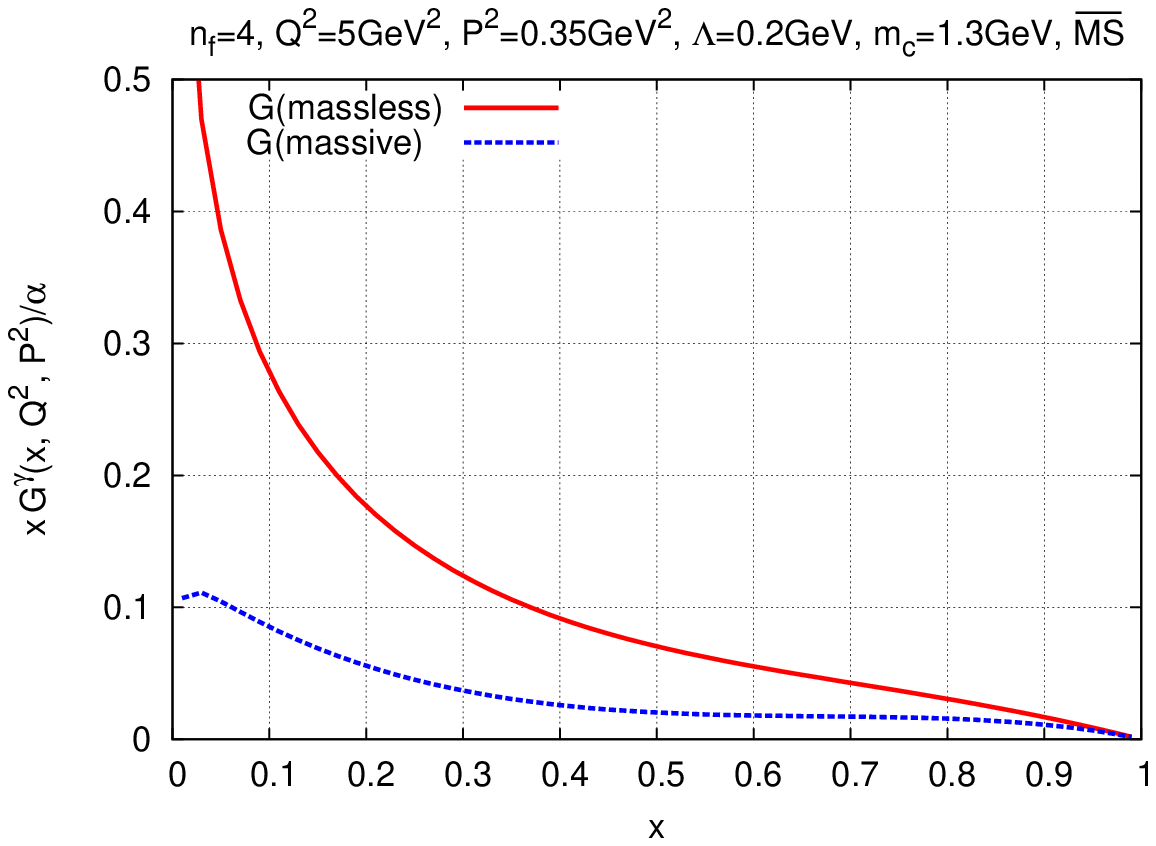} &
      \includegraphics[scale=\SCALE]{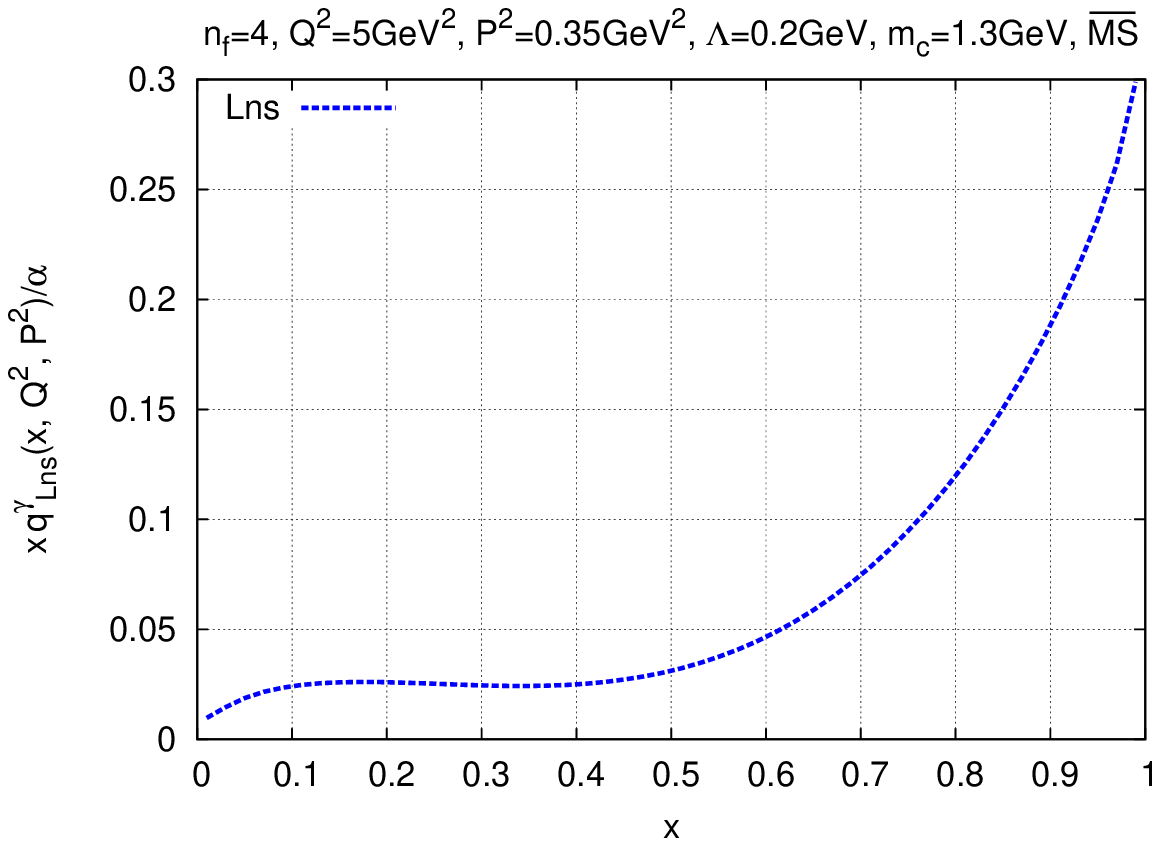} \\
      \hspace{\OFFSET} (c) & \hspace{\OFFSET} (d)
    \end{tabular}
    \caption{%
      Parton distributions in the photon in $\overline{\rm MS}$ scheme
      for $n_f=4$, $Q^2=5$GeV$^2$, $P^2=0.35$GeV$^2$ with $m_c=1.3$GeV
      and $\Lambda=0.2$GeV:
      (a) $x q_{Ls}^\gamma(x,Q^2,P^2)|_{\overline{\rm MS}}$ and
          $x q_{Ls}^\gamma(x,Q^2,P^2)|_{\text{light,}\overline{\rm MS}}$;
      (b) $x q_{H}^\gamma(x,Q^2,P^2)|_{\overline{\rm MS}}$ and
          $x q_{H}^\gamma(x,Q^2,P^2)|_{\text{light,}\overline{\rm MS}}$;
      (c) $x G^\gamma(x,Q^2,P^2)|_{\overline{\rm MS}}$ and
          $x G^\gamma(x,Q^2,P^2)|_{\text{light,}\overline{\rm MS}}$;
      (d) $x q_{Lns}^\gamma(x,Q^2,P^2)|_{\overline{\rm MS}}$.
    }
    \label{fig_MSbar_Q2=5}
  \end{center}
\end{figure}

\begin{figure}
  \begin{center}
    \def\SCALE{0.65}
    \def\OFFSET{27pt}
    \begin{tabular}{cc}
      \includegraphics[scale=\SCALE]{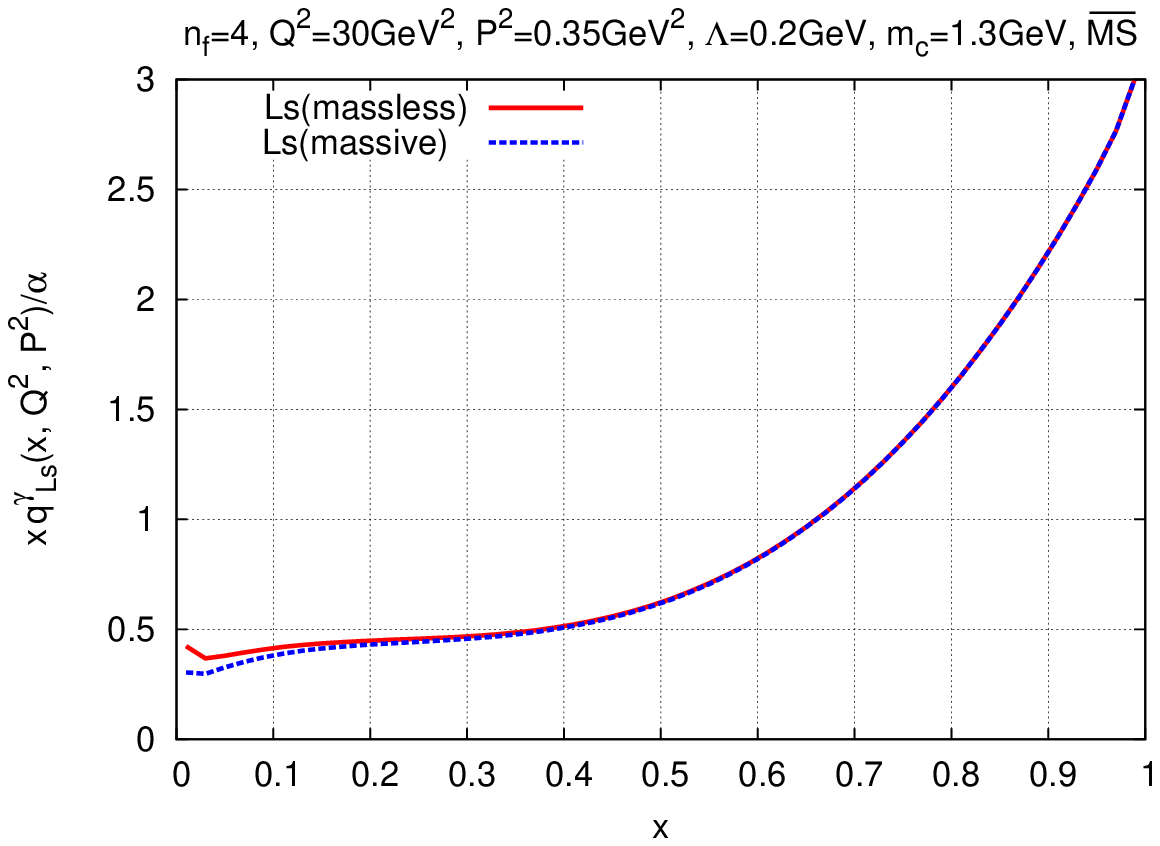} &
      \includegraphics[scale=\SCALE]{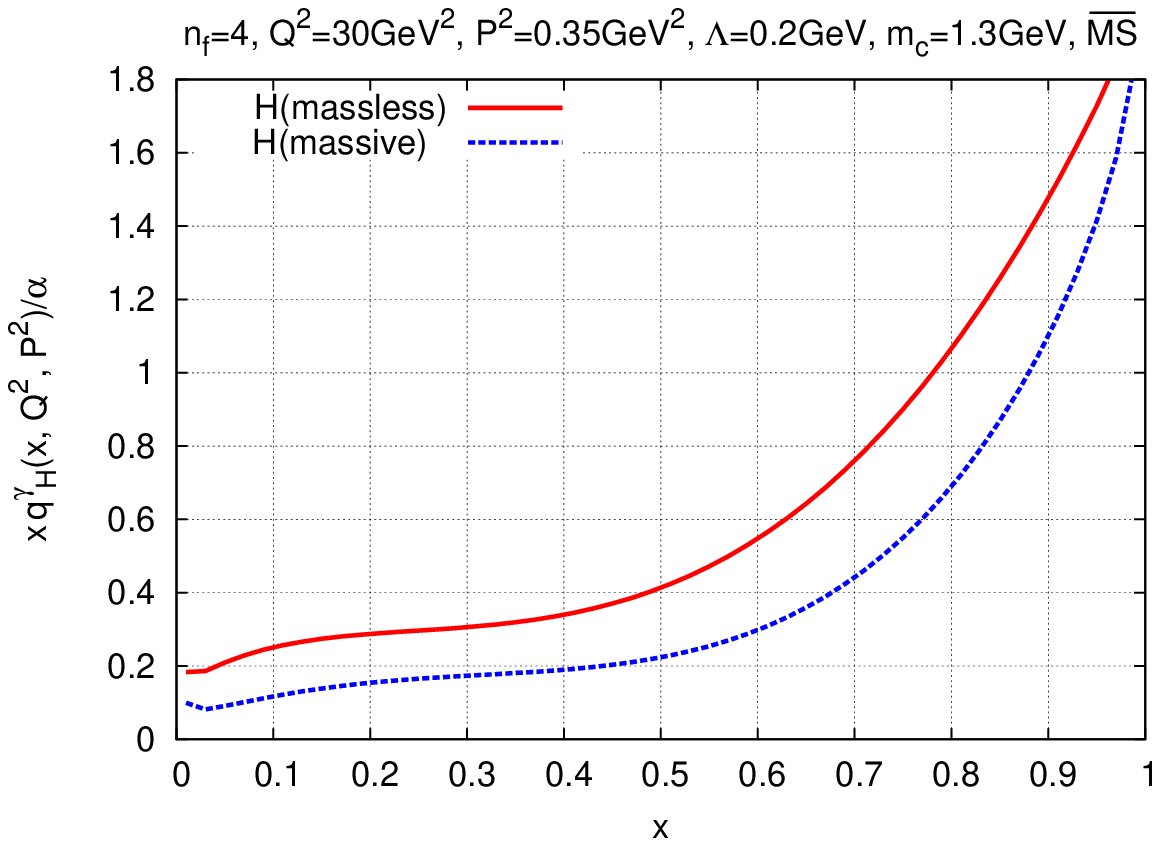} \\
      \hspace{\OFFSET} (a) & \hspace{\OFFSET} (b) \\
      \includegraphics[scale=\SCALE]{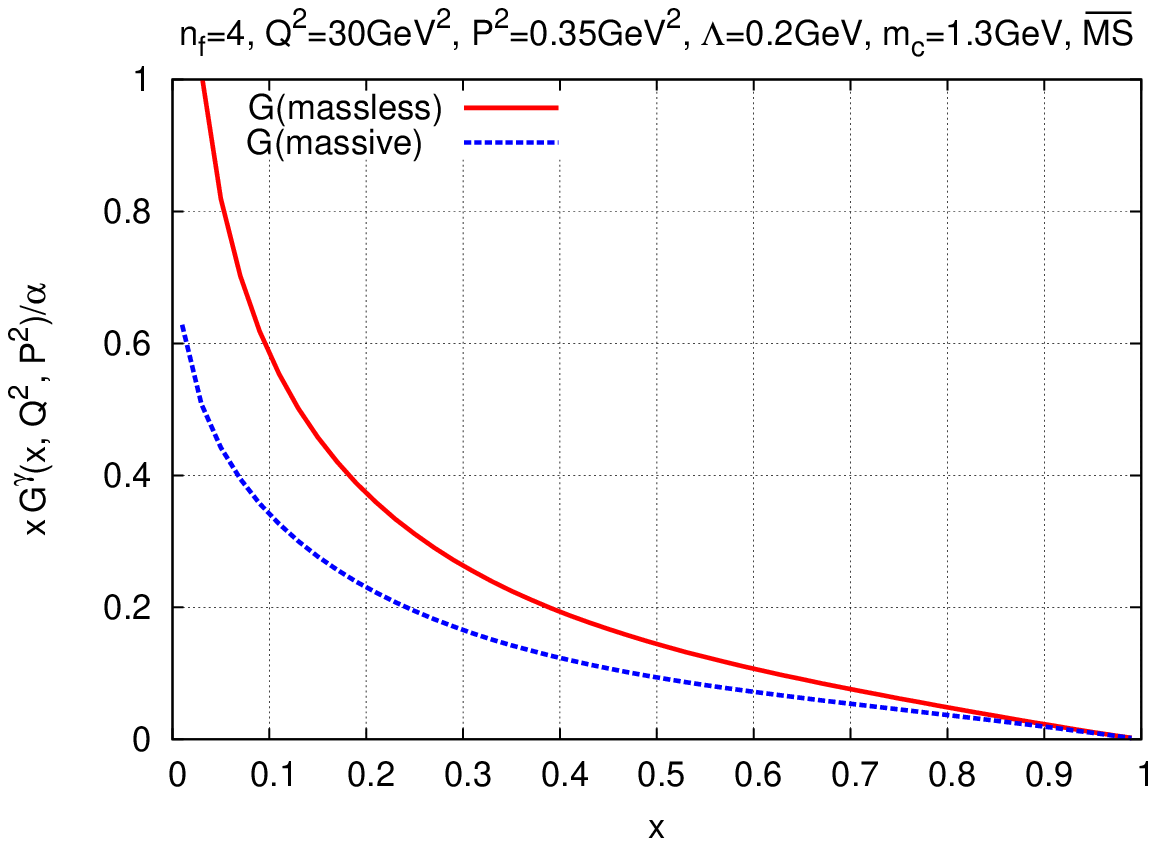} &
      \includegraphics[scale=\SCALE]{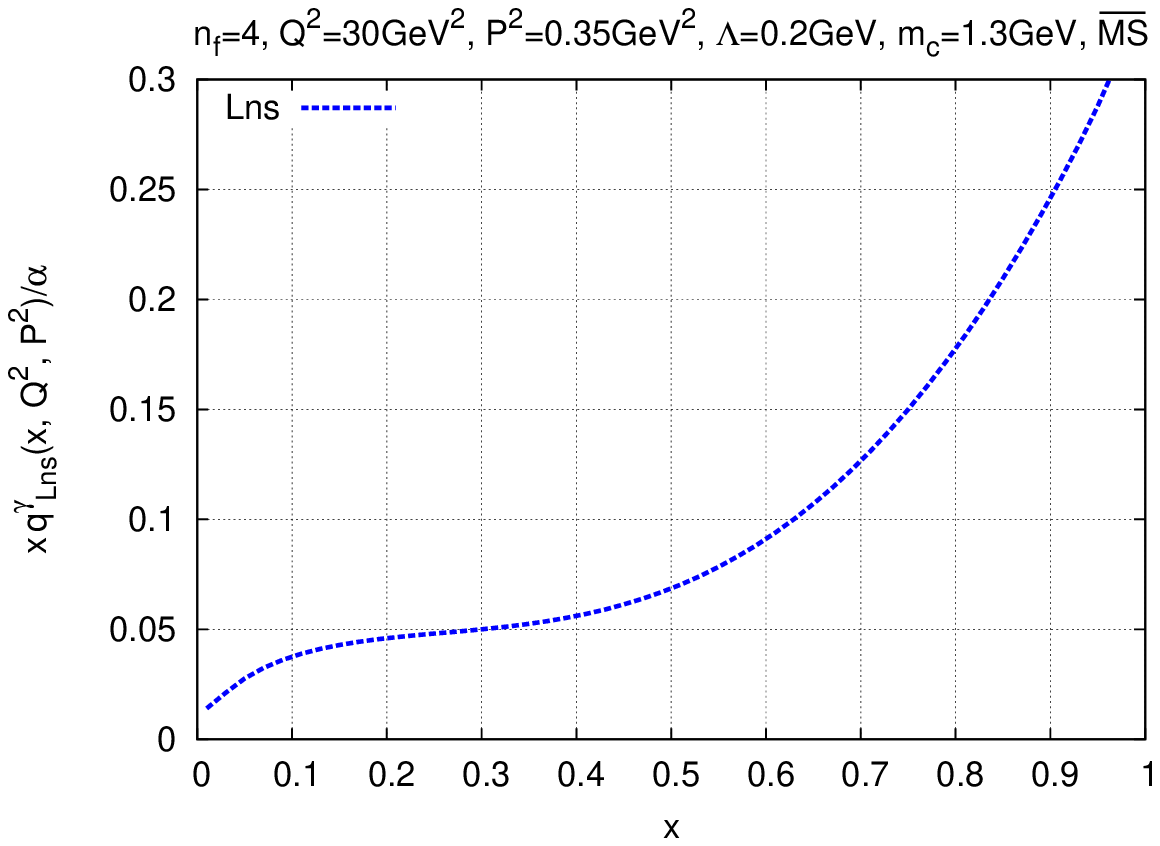} \\
      \hspace{\OFFSET} (c) & \hspace{\OFFSET} (d)
    \end{tabular}
    \caption{%
     Parton distributions in the photon in $\overline{\rm MS}$ scheme
      for $Q^2=30$GeV$^2$. The other parameters are the same as in Fig.~\ref{fig_MSbar_Q2=5}.
    }
    \label{fig_MSbar_Q2=30}
  \end{center}
\end{figure}

The values of $Q^2$ and $P^2$ for the case (i) correspond to those
of the PLUTO experiment~\cite{PLUTO}.
 We plot the parton distribution functions
in \MSbar{} scheme in Fig.~\ref{fig_MSbar_Q2=5} for the case (i)
and in Fig.~\ref{fig_MSbar_Q2=30} for the case (ii):
(a) the light ``flavor-singlet'' quark distribution~$xq^{\gamma}_{Ls}(x,Q^2,P^2)|_{\MSbar}$,
(b) the heavy (charm) quark distribution~$xq^{\gamma}_{H}(x,Q^2,P^2)|_{\MSbar}$,
(c) the gluon distribution~$xG^{\gamma}(x,Q^2,P^2)|_{\MSbar}$ and
(d) the light ``flavor-nonsinglet'' quark distribution~$xq^{\gamma}_{Lns}(x,Q^2,P^2)|_{\MSbar}$.
In order to see the heavy quark effects, we plot, in addition,
the parton distributions $xq^{\gamma}_{Ls}|_{\text{light},\MSbar}$,
$xq^{\gamma}_{H}|_{\text{light},\MSbar}$ and $xG^{\gamma}|_{\text{light},\MSbar}$
which are obtained when $c$-quark is also set to be massless.
Actually, we get these distributions by setting $\Delta {\widetilde A}^n_H \to 0$
in Eq.~\eqref{DefDeltaAH} and by inserting the ``new'' row vector ${\widetilde {\Av}}_n^{(1)}$
into the expression of Eq.~\eqref{qgammaONE}. See also Appendix~\ref{relation_massless}.
Since the light ``flavor-nonsinglet'' quark decouples to the other partons, we have
$xq^{\gamma}_{Lns}|_{\text{light}}=xq^{\gamma}_{Lns}$.

We observe from Fig.~\ref{fig_MSbar_Q2=5} (a)-(c) and Fig.~\ref{fig_MSbar_Q2=30} (a)-(c) that the $c$-quark mass has
rather large effects on the $c$-quark distribution $xq^{\gamma}_{H}|_{\MSbar}$
and the gluon distribution $xG^{\gamma}|_{\MSbar}$,
while it has negligible effects on the light ``flavor-singlet'' quark distribution~%
$xq^{\gamma}_{Ls}|_{\MSbar}$.
The difference between $xq^{\gamma}_{Ls}|_{\MSbar}$ and $xq^{\gamma}_{Ls}|_{\text{light},\MSbar}$
(and also between $xq^{\gamma}_{H}|_{\MSbar}$ and $xq^{\gamma}_{H}|_{\text{light},\MSbar}$
 and between $xG^{\gamma}|_{\MSbar}$ and $xG^{\gamma}|_{\text{light},\MSbar}$)
is due to
the appearance of $\Delta {\widetilde A}^n_H$ in Eq.~\eqref{DefDeltaAH},
which is negative and larger in magnitude for smaller $n$.
The negative $\Delta {\widetilde A}^n_H$ means that once $c$-quark has mass,
the $c$-partons are less produced from the target photon.
Actually it has the same effect as reducing the evolution of $c$-quark distribution
to the range of $m^2$ to $Q^2$ instead of $P^2$ to $Q^2$.
Indeed we find from Eq.~(\ref{qgammaONE}),
\bea
\left(q^{\gamma}_{Ls,n}-q^{\gamma}_{Ls,n}|_{\text{light}}\right)/\left(\frac{\alpha}{8\pi\beta_0}\right)
&=&
\left(1-\frac{1}{n_f}\right)2\beta_0\Delta {\widetilde A}^n_H
\bigg\{ -(r)^{d_\psi^n}+ \frac{\gamma_{\psi\psi}^{(0),n}-\lambda_-^n}{\lambda_+^n-\lambda_-^n}
(r)^{d_+^n}+ \frac{\gamma_{\psi\psi}^{(0),n}-\lambda_+^n}{\lambda_-^n-\lambda_+^n}
(r)^{d_-^n}\biggr\}~, \label{Deltaqls}
\\
\left(q^{\gamma}_{H,n}-q^{\gamma}_{H,n}|_{\text{light}}\right)/\left(\frac{\alpha}{8\pi\beta_0}\right)
&=&
2\beta_0\Delta {\widetilde A}^n_H
\biggl\{ \left(1-\frac{1}{n_f}\right)(r)^{d_\psi^n}+ \frac{1}{n_f}\frac{\gamma_{\psi\psi}^{(0),n}-\lambda_-^n}{\lambda_+^n-\lambda_-^n}
(r)^{d_+^n}+ \frac{1}{n_f}\frac{\gamma_{\psi\psi}^{(0),n}-\lambda_+^n}{\lambda_-^n-\lambda_+^n}
(r)^{d_-^n}\biggr\}~,\nn\\ \label{DeltaqH}
\eea
where $r=\frac{\alpha_s(Q^2)}{\alpha_s(P^2)}$. Unless $n$ is a small integer, we see
$\gamma_{\psi\psi}^{(0),n}\approx \lambda_-^n$ and $d_\psi^n \approx d_-^n$.
Therefore the sum in the curly brackets of Eq.~(\ref{Deltaqls}) diminishes,
which means that the effects of heavy quark on $xq^{\gamma}_{Ls}$
are extremely small. See Fig.~\ref{fig_MSbar_Q2=5} (a) and
Fig.~\ref{fig_MSbar_Q2=30} (a).
On the other hand, the sum in the curly brackets of Eq.~(\ref{DeltaqH})
is expressed approximately as $(r)^{d_\psi^n}$ for $n$ not being a small integer.
The ratio of $(q^{\gamma}_{H,n}-q^{\gamma}_{H,n}|_{\text{light}})$ to the leading order $q^{\gamma (0)}_{H,n}$
is proportional to the product of $\alpha_s(Q^2)$ and $(r)^{d_\psi^n}$.
The values of $r$ and $\alpha_s(Q^2)$ are 0.463 and 0.237,
respectively, for the case (i) and 0.351 and 0.180 for the case (ii).
The ratio is not small and Fig.~\ref{fig_MSbar_Q2=5} (b) and Fig.~\ref{fig_MSbar_Q2=30} (b) show
the large reduction of $xq^{\gamma}_{H}|_{\MSbar}$ from
$xq^{\gamma}_{H}|_{\text{light},\MSbar}$, especially, for the case (i).

The gluons do not couple to the photon directly and they are produced
from the target photon through quarks.
Therefore, the leading contribution to the gluon distribution $xG^{\gamma}|_{\MSbar}$ is essentially
of order $\alpha$ and it is very small in absolute value except
in the small $x$ region.
The $c$-quark mass effects appear in $xG^{\gamma}|_{\MSbar}$ in the NLO
(the order $\alpha$) and are enhanced by the charge factor $e_H=2/3$ (see Figs.~\ref{fig_MSbar_Q2=5} (c)
and~\ref{fig_MSbar_Q2=30} (c)).
The departure of $xq^{\gamma}_{Ls}|_{\MSbar}$ from
$xq^{\gamma}_{Ls}|_{\text{light},\MSbar}$ at small $x$ is
related to the behaviors of the gluon distributions $xG^{\gamma}|_{\MSbar}$
and $xG^{\gamma}|_{\text{light},\MSbar}$.
As $x \to 0$, the both gluon distributions grow while their difference becomes larger.

Figs.~\ref{fig_MSbar_Q2=5} (d) and~\ref{fig_MSbar_Q2=30} (d) show the light ``flavor-nonsinglet'' quark distribution~%
$xq^{\gamma}_{Lns}|_{\MSbar}$.
Comparing with the graphs of $xq^{\gamma}_{Ls}|_{\MSbar}$, it is very small
in absolute value.
This is due to the fact that $xq^{\gamma}_{Lns}|_{\MSbar}$ has the charge factor
$(\langle e^4 \rangle_L-(\langle e^2 \rangle_L)^2 )$
which is a very small number (2/81 for $n_f=4$).
An examination of Fig.~\ref{fig_MSbar_Q2=5} (b), (c) and
Fig.~\ref{fig_MSbar_Q2=30} (b), (c) show that
with larger $Q^2$, the $c$-quark mass effects become smaller.
When $Q^2$ gets still larger, we may need to consider the $b$-quark mass effects with taking $n_f=5$,
but $b$-quark has milder effects than $c$-quark because of its charge factor.

We see from Fig.~\ref{fig_MSbar_Q2=5} (a), (b), (d) and Fig.~\ref{fig_MSbar_Q2=30} (a), (b), (d)
that the quark distributions $xq^{\gamma}_{Ls}|_{\MSbar}$,
$xq^{\gamma}_{H}|_{\MSbar}$ and $xq^{\gamma}_{Lns}|_{\MSbar}$ diverge as $x \rightarrow 1$.
This is due to the NLO contributions to the quark parton distributions in ${\MSbar}$ scheme.
The behaviors of parton distributions near~$x=1$
are governed by the large-$n$ limit of those moments. In the leading order,
parton distributions are factorization-scheme independent.
For large $n$, the moments of the LO quark distributions,
$q^{\gamma (0)}_{Ls}$, $q^{\gamma (0)}_{H}$ and $q^{\gamma (0)}_{Lns}$, behave as
$1/(n\,\ln n)$. Thus, in $x$ space, these LO quark distributions vanish
for $x \rightarrow 1$ as $[-1/\ln(1-x)]$.
On the other hand, the moments of the NLO quark distributions in
$\overline {\rm MS}$ scheme,
$q^{\gamma (1)}_{Ls}\vert_{\overline {\rm MS}}$, $q^{\gamma (1)}_{H}\vert_{\overline {\rm MS}}$
and $q^{\gamma (1)}_{Lns}\vert_{\overline {\rm MS}}$, behave
in large-$n$ limit as $(\ln n)/n$. Therefore, in $x$ space, the (LO+NLO)
quark distributions in ${\overline {\rm MS}}$ scheme
{\it positively} diverge as $[-\ln(1-x)]$ for $x\rightarrow 1$.
The moments of the LO and NLO gluon distributions, $G_n^{\gamma (0)}$ and $G_n^{\gamma (1)}$,
behave for large $n$ as $1/(n\,\ln n)^2$ and $1/n^2$, respectively, and thus, in $x$ space,
the (LO + NLO) curve of the gluon distribution
(both $G^{\gamma}|_{\MSbar}$
and $G^{\gamma}|_{\text{light},\MSbar}$) vanishes as $(-\ln x)$ for $x\rightarrow 1$.

\begin{figure}
  \begin{center}
    \def\SCALE{0.65}
    \def\OFFSET{27pt}
    \begin{tabular}{cc}
      \includegraphics[scale=\SCALE]{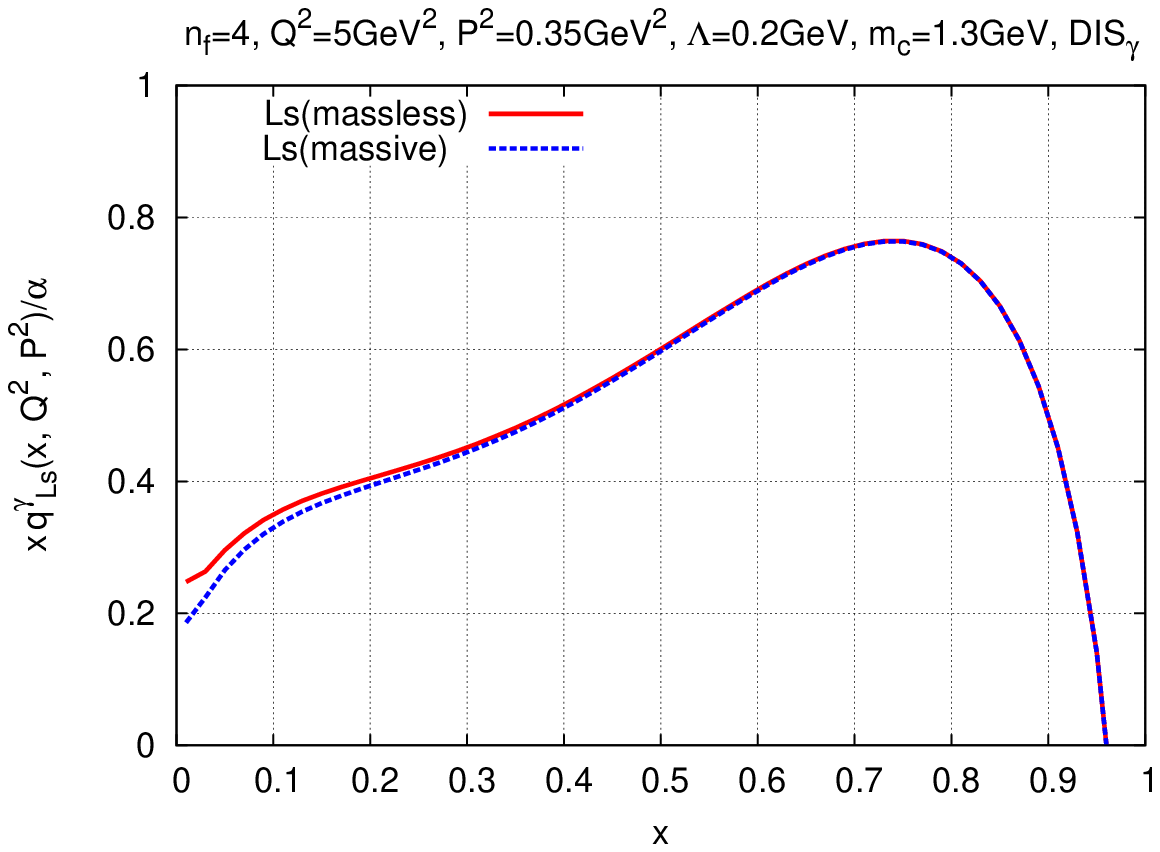} &
      \includegraphics[scale=\SCALE]{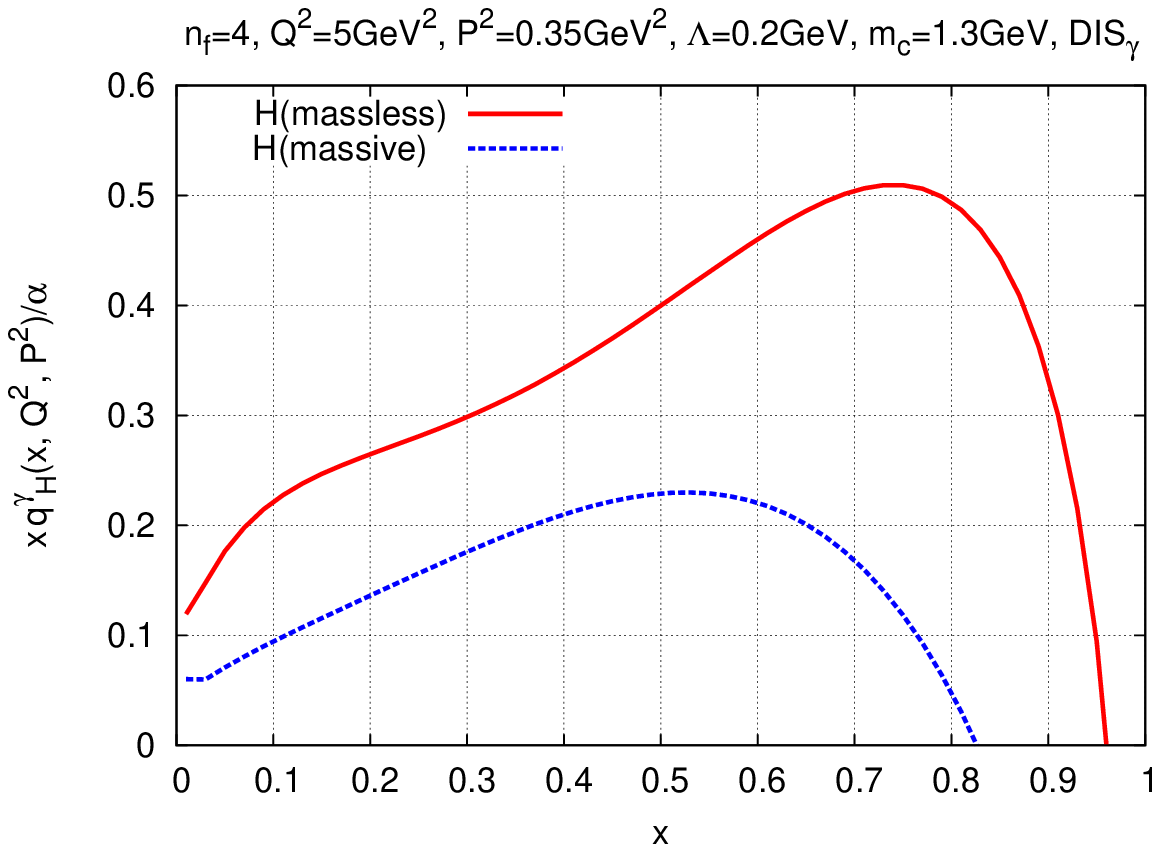} \\
      \hspace{\OFFSET} (a) & \hspace{\OFFSET} (b) \\
      \multicolumn{2}{c}{%
      \includegraphics[scale=\SCALE]{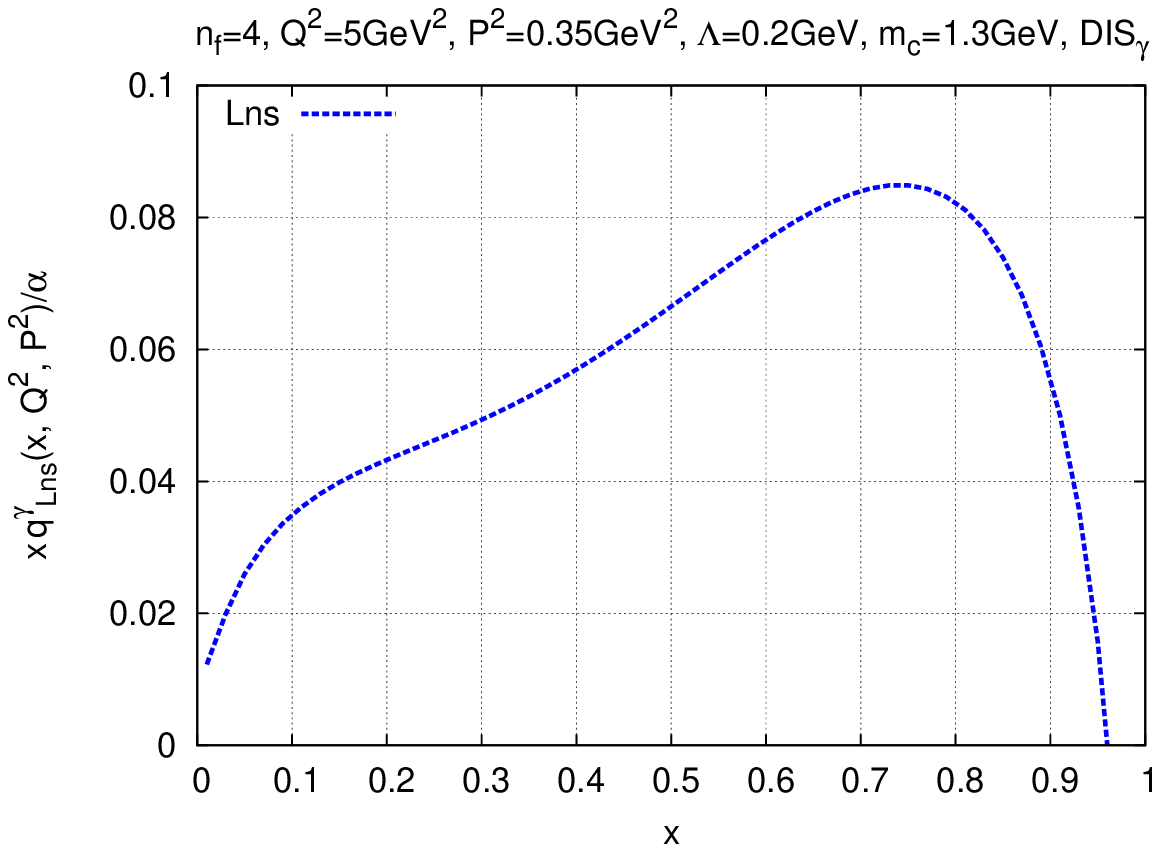}} \\
      \multicolumn{2}{c}{%
      \hspace{\OFFSET} (c)}
    \end{tabular}
    \caption{%
      Parton distributions in the photon in DIS$_\gamma$ scheme
      for $n_f=4$, $Q^2=5$GeV$^2$, $P^2=0.35$GeV$^2$ with $m_c=1.3$GeV
      and $\Lambda=0.2$GeV:
      (a) $x q_{Ls}^\gamma(x,Q^2,P^2)|_{\text{DIS}_\gamma}$ and
          $x q_{Ls}^\gamma(x,Q^2,P^2)|_{\text{light,DIS}_\gamma}$;
      (b) $x q_{H}^\gamma(x,Q^2,P^2)|_{\text{DIS}_\gamma}$ and
          $x q_{H}^\gamma(x,Q^2,P^2)|_{\text{light,DIS}_\gamma}$;
      (c) $x q_{Lns}^\gamma(x,Q^2,P^2)|_{\text{DIS}_\gamma}$.
    }
    \label{fig_DISg_Q2=5}
  \end{center}
\end{figure}

\begin{figure}
  \begin{center}
    \def\SCALE{0.65}
    \def\OFFSET{27pt}
    \begin{tabular}{cc}
      \includegraphics[scale=\SCALE]{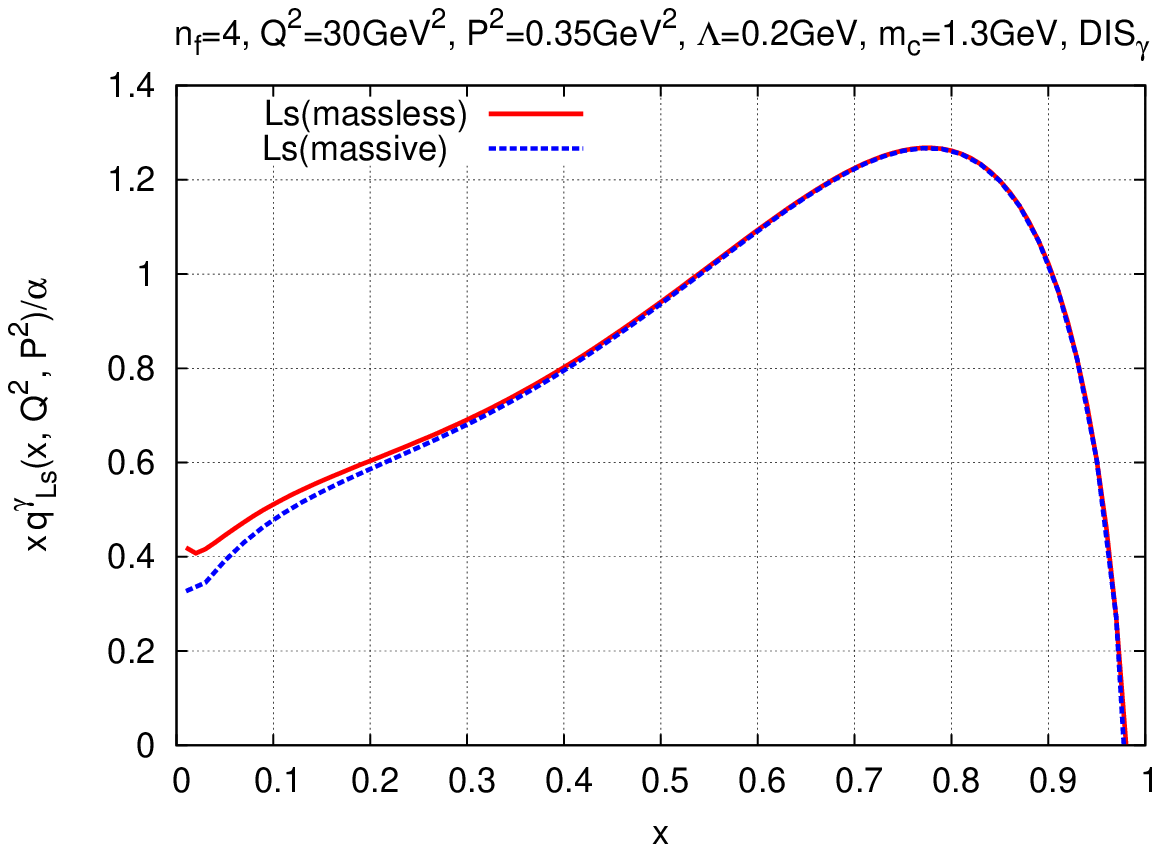} &
      \includegraphics[scale=\SCALE]{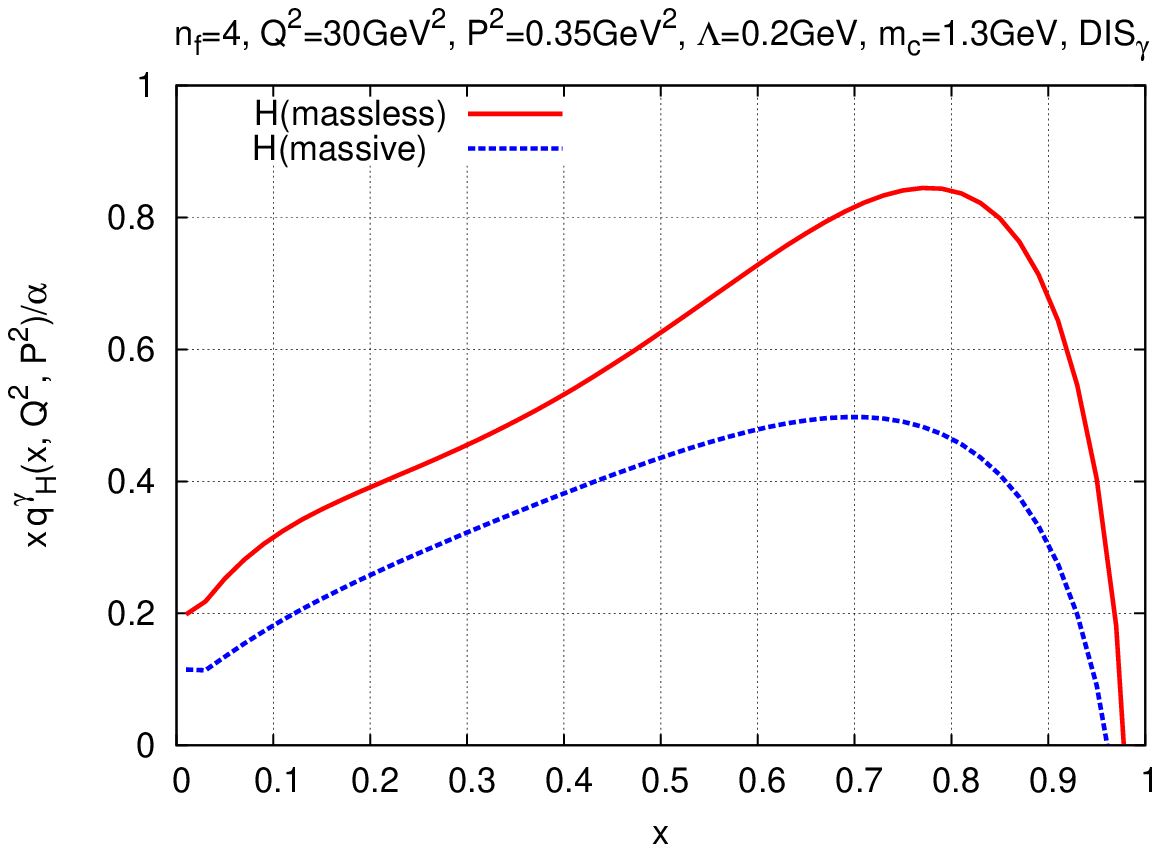} \\
      \hspace{\OFFSET} (a) & \hspace{\OFFSET} (b) \\
      \multicolumn{2}{c}{%
      \includegraphics[scale=\SCALE]{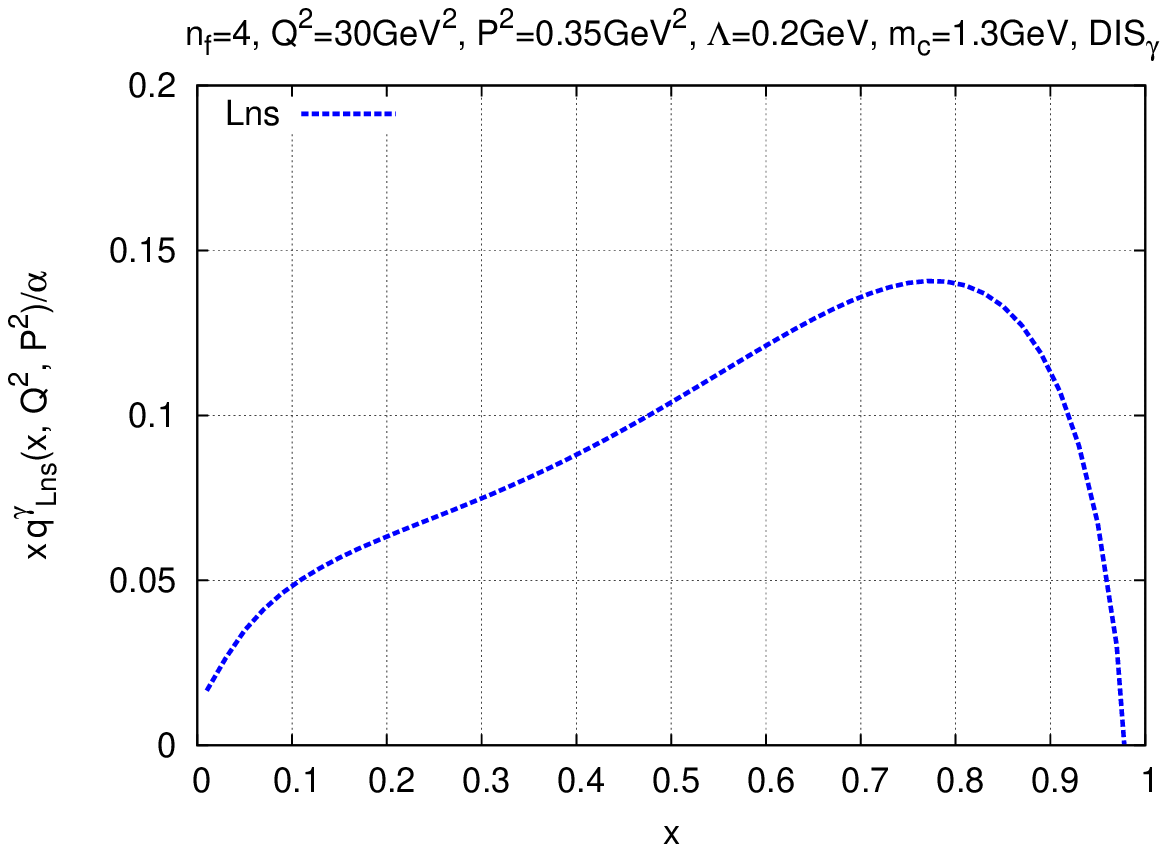}} \\
      \multicolumn{2}{c}{%
      \hspace{\OFFSET} (c)}
    \end{tabular}
    \caption{%
 Parton distributions in the photon in DIS$_\gamma$ scheme
      for $Q^2=30$GeV$^2$. The other parameters are the same as in
      Fig.~\ref{fig_DISg_Q2=5}.
    }
    \label{fig_DISg_Q2=30}
  \end{center}
\end{figure}

In \DISg{} scheme the photonic coefficient $C_2^\gamma$ is absorbed into the quark distributions.
In consequence, the (LO+NLO) quark distributions show different behaviors at large $x$ from those in \MSbar{} scheme.
Since $B_\gamma^{L,n}(=B_\gamma^{H,n})$ in Eq.~(\ref{CoefficientGammaNLO}) behaves as $(-4\ln n)/n$
for large $n$, the (LO+NLO) curves in $x$ space for
$q^{\gamma}_{Ls}\vert_{\DISg}$, $q^{\gamma}_{H}\vert_{\DISg}$ and $q^{\gamma }_{Lns}\vert_{\DISg}$
{\it negatively} diverge as $\ln(1-x)$ for $x\rightarrow 1$.
In fact, using Eqs.~\eqref{deltaqLs(1)}-\eqref{GDISg} and inverting the moments, we obtain
parton distributions in \DISg{} scheme up to the NLO, which are
plotted in Fig.~\ref{fig_DISg_Q2=5} and Fig.~\ref{fig_DISg_Q2=30}.
Again we have considered the two cases: (i) $Q^2=5\GeV^2$, $P^2=0.35\GeV^2$ and
(ii) $Q^2=30\GeV^2$, $P^2=0.35\GeV^2$. The other parameters are the same as before and
$c$-quark is taken to be heavy. We see from Fig.~\ref{fig_DISg_Q2=5} (a), (b), (c)
and Fig.~\ref{fig_DISg_Q2=30} (a), (b), (c) that the quark distributions $xq^{\gamma}_{Ls}|_{\DISg}$,
$xq^{\gamma}_{H}|_{\DISg}$ and $xq^{\gamma}_{Lns}|_{\DISg}$ become negative at large $x$.
We observe again that the mass of $c$-quark has
negligible effects on the light ``flavor-singlet'' quark distribution~
$xq^{\gamma}_{Ls}|_{\DISg}$ but
large effects on the $c$-quark distribution $xq^{\gamma}_{H}|_{\DISg}$.
When $Q^2$ gets larger, the heavy quark mass effects become smaller.
It is noted that if we take into account the charge factors, the following three ``renormalized'' distributions,
$x{\widetilde q}^{\gamma}_{Ls}|_{\text{light},\DISg}
  \equiv xq^{\gamma}_{Ls}|_{\text{light},{\DISg}} / \langle e^2 \rangle_L$,
$x{\widetilde q}^{\gamma}_{H}|_{\text{light},{\DISg}}
  \equiv xq^{\gamma}_{H}|_{\text{light},\DISg} / e_H^2$
and $x{\widetilde q}^{\gamma}_{Lns}|_{\DISg}
  \equiv xq^{\gamma}_{Lns}|_{\DISg} / \left( \langle e^4 \rangle_L-(\langle e^2 \rangle_L)^2 \right)$
overlap for almost the whole $x$ region except near $x=0$
(see Fig.~\ref{fig_DISg_Q2=5} (a), (b) and (c) and
Fig.~\ref{fig_DISg_Q2=30} (a), (b) and (c) ).

Finally the gluon distribution $xG^{\gamma}|_{\DISg}$ is the same as
$xG^{\gamma}|_{\MSbar}$.

\section{Conclusions}

We have studied the heavy quark mass effects on the parton
(light singlet, heavy quark, gluon, light nonsinglet)
distribution functions in the virtual photon up to the NLO in perturbative QCD.
Our calculation is based on DGLAP equation as well as on the OPE formalism
within the framework of the mass-independent renormalization group.
The heavy quark effect is included through the operator matrix element
for heavy quark operator and is evaluated by the heavy quark mass limit
($\Lambda^2 \ll P^2 \ll m^2$). In the language of the parton picture, the
heavy quark effects are arising from the initial condition for the heavy quark
distribution.
In fact, the leading-logarithimic term of our initial condition (\ref{deltaAH})
can be reproduced by solving
the boundary condition, $ q_H^\gamma(x,Q^2=m^2)=0$~\cite{Fontannaz,AFG},
in the leading order approximation.

The heavy quark mass effects tend to reduce
the values of parton distribution functions for
the light-singlet, the heavy-quark and the gluon distributions
except for the light nonsinglet distribution.
Especially the suppression for the heavy parton distribution for the up-type
quark is enhanced by the dependence of the charge factor $e_{i}=2/3$.
These behaviors are consistent with our previous work
on the virtual photon structure functions with the heavy quark
mass effects~\cite{KSUU2009}.
These results could be explained by the suppression
of the evolution range due to the mass of the heavy quark.
We have also studied the factorization-scheme dependence of our
parton distributions with heavy quark mass effects, especially for
two factorization schemes, $\overline{\rm MS}$ and DIS$_\gamma$.

In our formalism where we treat the contribution from the twist-2
operators to the parton distributions,
we have not taken into account the kinematical threshold effects which
manifest as the presence of the maximal values of the Bjorken variable.
We need some improvement in which the threshold effects are included.
We should also investigate the general kinematical region where $P^2$
and $m^2$ are of the same order. We also note that the general-mass
variable-flavor-number scheme (GM-VFNS), which has now become a popular
framework for the global analyses of parton distributions,
should be implemented in the present analysis which is under investigation.

Such improvements would help us to predict the photonic parton
distribution functions which could be measured at the future linear
collider ILC.
Especially, the application of our results to the up-type quark parton
distribution functions like the charm quark, the top quark
in the unpolarized virtual photon will be important phenomenologically.
The application of our formalism to the polarized photonic
parton distribution functions can be carried out and would turn out to
be relevant for the measurement of the polarized photonic PDFs at ILC.

%------ Appendix -------%
\appendix

\section{PDFs for the case of massless quarks}
\label{relation_massless}

In this paper we have considered the parton distributions in the virtual photon
for the case when the $n_f$-th flavor quark is heavy and the rest of $n_f -1$ flavor
quarks are light (i.e., massless) and we have derived the formulae for the moments
of the parton distributions, $q^{\gamma}_{Ls}(n,Q^2,P^2)$,
$q^{\gamma}_{H}(n,Q^2,P^2)$, $G^{\gamma}(n, Q^2, P^2) $ and $q^{\gamma}_{Lns}(n,Q^2,P^2)$
up to NLO, which are given in Eqs.~(\ref{Mofqls})-(\ref{Mofqlns}).
If the $n_f$-th flavor quark is also light, in other words, all the $n_f$
flavor quarks are light,
we obtain, instead, the parton distributions of
$q^{\gamma}_{Ls}(n,Q^2,P^2)|_{\mbox{\scriptsize light}}$,
$q^{\gamma}_{H}(n,Q^2,P^2)|_{\mbox{\scriptsize light}}$ and $G^{\gamma}(n, Q^2, P^2)|_{\mbox{\scriptsize light}}$
and $q^{\gamma}_{Lns}(n,Q^2,P^2)|_{\mbox{\scriptsize light}}$.
Here it is noted that since the light ``flavor nonsinglet'' quark does not couple to the heavy flavor,
we see $q^{\gamma}_{Lns}(n,Q^2,P^2)|_{\mbox{\scriptsize light}}=q^{\gamma}_{Lns}(n,Q^2,P^2)$.
When all the flavor quarks are light, we usually
treat the quark sector which consists of the flavor-singlet and nonsinglet combinations
defined as follows:
\be
q_S^\gamma\equiv \sum_{i=1}^{n_f}q^i~, \qquad q^\gamma_{NS}\equiv
\sum_{i=1}^{n_f} e^2_i \Bigl(q^{i} - \frac{q^{\gamma}_{S}}{n_f }\Bigr) .
\ee
Also we introduce the following quark-charge factors:
\be
\langle e^2 \rangle=\frac{1}{n_f}\sum_{i=1}^{n_f}e_i^2~,\qquad
\langle e^4 \rangle=\frac{1}{n_f}\sum_{i=1}^{n_f}e_i^4~. \label{ChargeFactorLight}
\ee
These parton distributions, $q_S^\gamma$, $q^\gamma_{NS}$ and
$G^{\gamma}|_{\mbox{\scriptsize light}}$, have been investigated in
Refs.~\cite{GRV1992a,MVV2002,USU2009}.
The quark parton distributions
$q^{\gamma}_{Ls}|_{\mbox{\scriptsize light}}$, $q^{\gamma}_{H}|_{\mbox{\scriptsize light}}$ and $q^{\gamma}_{Lns}$
are related to $q_S^\gamma$ and $q^\gamma_{NS}$. Indeed they
are expressed in terms of $q_S^\gamma$ and $q^\gamma_{NS}$ as follows:
\begin{eqnarray}
q^{\gamma}_{Ls}(n,Q^2,P^2) \big|_{\mbox{\scriptsize light}}
 &=& q^{\gamma}_{S}(n,Q^2,P^2) - q^{\gamma}_{H}(n,Q^2,P^2)
 \big|_{\mbox{\scriptsize light}}, \\
%---
 q^{\gamma}_{H}(n,Q^2,P^2) \big|_{\mbox{\scriptsize light}}
 &=& \frac{1}{n_f}
     \left [   q^{\gamma}_{S}(n,Q^2,P^2)
             + \frac{e_{H}^2 - \langle e^2 \rangle}
	            {\langle e^4 \rangle - \langle e^2 \rangle^2}~
		    q^{\gamma}_{NS}(n,Q^2,P^2)
     \right], \label{non-trivial-eq} \\
%---
q^{\gamma}_{Lns}(n,Q^2,P^2)
&=& \frac{n_f - 1}{n_f}
    \frac{\langle e^4 \rangle_{L} - \langle e^2 \rangle ^2_{L}}
         {\langle e^4 \rangle - \langle e^2 \rangle^2}~
    q^{\gamma}_{NS}(n,Q^2,P^2),
\end{eqnarray}
where the NLO expressions of $q^{\gamma}_{S}(n,Q^2,P^2)$ and
$q^{\gamma}_{NS}(n,Q^2,P^2)$ are given in Eqs.~(2.33), (2.35) and~(2.37)
of Ref.~\cite{USU2009}.
The transformation rule for $q^{\gamma}_{S}$ and $q^{\gamma}_{NS}$
from the $\rm{\overline{MS}}$ scheme to the ${\rm DIS}_{\gamma}$ scheme
are given in Eqs.~(3.33) and~(3.34), respectively, of the same reference.

%------ End of Appendix -------%

%------ References -------%
%\bibliography{apssamp}

\end{document}